\documentclass[
showpacs,
preprintnumbers,
 amsmath,amssymb,
 aps,
floatfix
]{revtex4-1}

\usepackage{verbatim}
\usepackage[export]{adjustbox}
\usepackage{float}
\usepackage{subcaption}

\usepackage{graphicx}% Include figure files
\usepackage{dcolumn}% Align table columns on decimal point
\usepackage{bm}% bold math
\usepackage{hyperref}% add hypertext capabilities
\usepackage[mathlines]{lineno}% Enable numbering of text and display math
\newcommand{\Tr}[1]{\mathrm{Tr}\left[#1\right]}
\newcommand{\expo}[1]{\mathrm{e}^{#1}}
\newcommand{\bra}[1]{\langle #1 \vert}
\newcommand{\ket}[1]{\vert #1 \rangle}
\newcommand{\braket}[2]{\langle #1 \vert #2 \rangle}

\begin{document}

%\preprint{APS/123-QED}

\title{Finite-temperature degenerate perturbation theory for bosons in optical lattices}% Force line breaks with \\

\author{Felipe Taha Sant'Ana}
%\author{Axel Pelster}%
 \email{felipe.taha@usp.br}
\affiliation{%
 S\~ao Carlos Institute of Physics, University of S\~ao Paulo \\ 13566-590 S\~ao Carlos, São Paulo, Brazil
}%


\author{Axel Pelster}
 \email{axel.pelster@physik.uni-kl.de}
\affiliation{
 Physics Department and Research Center OPTIMAS, Technische Universit\"at Kaiserslautern \\ 67663 Kaiserslautern, Germany% with \\
}%
%\affiliation{
 %Third institution, the second for Charlie Author
%}%
\author{Francisco Ednilson Alves dos Santos}
 \email{santos@ufscar.br}
\affiliation{%
 Physics Department, Federal University of S\~ao Carlos \\ 13565-905 S\~ao Carlos, São Paulo, Brazil
}%


\date{\today}% It is always \today, today,

\begin{abstract}
Bosonic atoms confined in optical lattices can exist in two different phases, Mott insulator and superfluid, depending on the strength of the system parameters, such as the on-site interaction between particles and the hopping parameter. This work is motivated by the fact that nondegenerate perturbation theory applied to the mean-field approximation of the Bose-Hubbard Hamiltonian at both zero and finite temperature fails to give consistent results in the vicinity of the Mott insulator-superfluid phase transition, e.g., the order parameter calculated via nondegenerate perturbation theory reveals an unphysical behavior between neighboring Mott lobes, which is an explicit consequence of degeneracy problems that artificially arise from such a treatment. Therefore, in order to fix this problem, we propose a finite-temperature degenerate perturbation theory approach based on a projection operator formalism which ends up solving such degeneracy problems in order to obtain physically consistent results for the order parameter near the phase transition.
%\begin{description}
%\end{description}
\end{abstract}

\pacs{67.85.−d, 42.50.−p, 32.80.Pj, 03.75.Lm}% PACS, the Physics and Astronomy
\maketitle

%\tableofcontents

\section{\label{sec1}Introduction}

Optical lattices are laser arrangements which enable a spatially periodic  trapping of atoms due to the interaction between the external electric field and the induced dipole moment of the atoms \cite{jaksch,bloch,pethick,pitaevskii}. A gas composed of bosonic atoms in an optical lattice can be described by the Bose-Hubbard model \cite{ueda,lewenstein}, which has three main parameters: the on-site interaction parameter, the hopping parameter, and the chemical potential. Depending on the magnitude of the parameters, the system can realize two different phases, the Mott insulator or the superfluid phase \cite{fisher,greiner1,greiner2,widera,folling,gunter,ospelkaus,lewenstein2,gerbier}. If the on-site interaction parameter is much larger than the hopping parameter, the system is in the Mott insulator (MI) phase. This phase is characterized by a strong localization of the atoms. By decreasing the amplitude of the periodic potential, so that the hopping parameter becomes much larger than the atom-atom interaction parameter, the system undergoes a phase transition to a superfluid (SF) phase, where the atoms are delocalized. Such differences in the localization of bosons make it possible to measure the phase the system is currently in through time-of-flight experiments \cite{greiner1,hoff}. The MI-SF transition can happen even at zero-temperature, driven by quantum mechanical fluctuations, thus characterizing a quantum phase transition \cite{sachdev}.

The main difficulty in solving the Bose-Hubbard Hamiltonian is the nonlocality of the hopping term. Thus, a common path for obtaining a first approximation of the MI-SF quantum phase diagram is a mean-field calculation, which approximates the Bose-Hubbard Hamiltonian by a sum of local Hamiltonians \cite{fisher}. Following this simplification, Rayleigh-Schr\"odinger perturbation theory (RSPT) is typically used for obtaining the mean-field phase diagram at zero temperature \cite{sachdev}. However, there are problems that arise from RSPT, since it does not properly deal with the degeneracies that occur between two consecutive Mott lobes. One of such RSPT problems concerns the calculation of the condensate order parameter which falsely vanishes between consecutive Mott lobes \cite{melo,martin}.

Also, other methods have been suggested in order to improve the mean-field quantum phase diagram for bosons in optical lattices, such as in Ref.~\cite{santos}, which uses a variational method and the field-theoretic concept of the effective potential. Also, the MI-SF phase transition at arbitrary temperature was investigated in Ref. \cite{bradlyn} using an effective action approach. Furthermore, in Refs. \cite{grass1,grass2} an effective action approach was derived for the Bose-Hubbard model within the Schwinger-Keldysh formalism in order to handle time-dependent problems at finite temperature. Likewise, \cite{melo} implemented a nearly degenerate perturbation theory for the zero-temperature case, which led to better results for the order parameter (OP) when compared to those from the RSPT calculations. More recently, Brillouin-Wigner perturbation theory was applied in order to correct such degeneracy-generated unphysical results at zero-temperature \cite{martin}. It turns out that nondegenerate finite-temperature perturbation theory, as applied in Refs. \cite{bradlyn,ednilson}, also presents degeneracy problems similar to RSPT. Indeed, RSPT is equivalent to the usual finite-temperature perturbation theory in the zero-temperature limit. Therefore, degeneracy-related problems are also expected to appear at low enough temperatures.

The present work is concerned with correcting the degeneracy problem that artificially arises from such perturbative approaches regarding a system composed of bosonic atoms confined in an optical lattice at finite temperature. Starting from the mean-field approximation for the Bose-Hubbard Hamiltonian and considering the Landau expansion for the order parameter in the vicinity of the MI-SF phase transition, we perform a perturbation theory in imaginary time. In addition, in order to fix the degeneracy problem, we introduce a projection operator formalism for the finite-temperature system. The main idea of this degenerate approach is to separate the Hilbert subspace in which the degeneracies are contained from its complement. This system is then exactly diagonalized inside the degenerate subspace, while the effects of the interaction between the two subspaces are taken into account perturbatively. Such a procedure corrects the degeneracy problem and leads to physically consistent results for the condensate density.

In Sec. \ref{sec2}, we introduce the mean-field approximation for the Bose-Hubbard Hamiltonian in order to get rid of its nonlocality, which transforms the originial Bose-Hubbard Hamiltonian into a sum of local Hamiltonians, thus allowing us to work with separated lattice sites. In Sec. \ref{landau}, we briefly discuss the Landau theory for second-order phase transitions, which enables us to expand the system free energy as a power series of the OP in the vicinity of the MI-SF phase transition. Furthermore, the Landau theory gives us the equation for the phase boundary. In Sec. \ref{PT}, we apply the imaginary-time-dependent nondegenerate perturbation theory (NDPT) considering the system in the vicinity of the phase transition. In this way, we treat the OP perturbatively in order to get expressions for the Landau coefficients and, consequently, obtain the phase boundary as well as the OP, close to the phase boundary. Following the calculation of the Landau coefficients, Sec. \ref{inconsistencies} exposes the unphysical behavior of the OP between two consecutive Mott lobes, which is an explicit consequence of the degeneracies that are not adequately handled within the framework of NDPT at low enough temperatures. This inconsistency in the theory is the motivation for our proposed degenerate approach, which is worked out in detail in Sec. \ref{sec3}. In Sec. \ref{CD}, we evaluate the condensate densities for different temperatures and values of system parameters making use of our proposed finite-temperature degenerate perturbation theory (FTDPT). In Sec. \ref{comparison}, we turn our attention to a region between two consecutive Mott lobes in the phase diagram, where the superfluid clearly dominates and also a region where the NDPT fails at very low temperatures. We compare the results of the NDPT and the FTDPT to conclude that our degenerate method corrects all inconsistencies. Finally, we calculate the particle densities in Sec. \ref{density} for different temperatures and describe the existence of a melting of the wedding cake like structure.

\section{\label{sec2}Mean-Field Approximation}
The description of spinless bosonic atoms confined in an optical lattice is given by the Bose-Hubbard Hamiltonian,
\begin{equation}\label{BH}
\hat{H}_{BH}=\frac{U}{2}\sum_{i}\hat{a}_{i}^{\dagger}\hat{a}_{i}^{\dagger}\hat{a}_{i}\hat{a}_{i}-t\sum_{\langle i,j\rangle}\hat{a}_{i}^{\dagger}\hat{a}_{j}-\mu\sum_{i}\hat{a}_{i}^{\dagger}\hat{a}_{i}.
\end{equation}
The respective parameters are the following: $t$ represents the hopping parameter, $U$ stands for the on-site interaction parameter describing the interaction between particles, and $\mu$ denotes the chemical potential within a grand-canonical description. Furthermore, $\hat{a}^\dagger_i$ and $\hat{a}_i$ are the usual bosonic creation and annihilation operators at site $i$, respectively. Note that in this model only nearest neighbor hopping is allowed and this restriction is depicted by $\langle i,j \rangle$.

Due to the nonlocal character of the hopping parameter, a standard mean-field approximation is usually the simplest way to solve a problem of this kind. The fundamental concept behind such an approach is to approximate the nonlocal hopping term by a local one. This procedure results in the mean-field Hamiltonian \cite{melo,ednilson,santos,fisher},
\begin{equation}
	\label{mfH}
\hat{H}_{MF}=\frac{U}{2}\sum_i \left(\hat{n}_i^{2}-\hat{n}_i\right) -\sum_i \mu\hat{n}_i -tz\sum_i\left(\Psi^{*}\hat{a}_i+\Psi\hat{a}_i^{\dagger}-\Psi^{*}\Psi\right) ,
\end{equation}
where $z$ denotes the number of nearest neighbors, $\Psi\equiv\langle\hat{a}_{i}\rangle$ and $\hat{n}_i\equiv\hat{a}^{\dagger}_i\hat{a}_i$ is the number operator. Since (\ref{mfH}) is a sum of local Hamiltonians, we restrict ourselves in the following to the one lattice site Hamiltonian,
\begin{equation}
\label{H_one}
\hat{H}=\frac{U}{2}\left(\hat{n}^{2}-\hat{n}\right) -\mu\hat{n} -tz\left(\Psi^{*}\hat{a}+\Psi\hat{a}^{\dagger}-\Psi^{*}\Psi\right).
\end{equation}

\subsection{\label{landau}Landau theory}
Landau argued that the free energy can be written as a polynomial function of the order parameter in the vicinity of a phase transition \cite{landau}, 
\begin{equation}\label{landauexpansion}
    \mathcal{F} \left( \Psi^*,\Psi\right) = a_0 +  a_2 |\Psi|^2 + a_4 |\Psi|^4 + \cdots .
\end{equation}

Since $\mathcal{F}$ is considered to be an analytic function of $\Psi$ and since the Bose-Hubbard Hamiltonian described by (\ref{BH}) possesses a global $U(1)$ phase invariance, $a_n$ will not vanish only for even values of $n$. Therefore, for small values of $|\Psi|$, we can consider only the lowest-order terms in (\ref{landauexpansion}), i.e., $a_0$, $a_2$, and $a_4$ as nonvanishing. For $a_4>0$, a second-order phase transition may occur. This originates from the fact that, if $a_2>0$, the only solution of $\partial \mathcal{F}/\partial \Psi = 0$ is $\Psi = 0$, thus corresponding to the MI symmetric phase, while if $a_2<0$, the effective potential $\mathcal{F}$  has infinitely many minima with $|\Psi|\neq 0$ which differ only in the phase of $\Psi$ and corresponds to the SF phase \cite{ednilson}. Thus we conclude that the condition $a_2=0$ defines the boundary between the two phases.

\subsection{\label{PT}Nondegenerate perturbation theory}
As mentioned before, the transition from Mott insulator to superfluid is followed by a symmetry breaking and can be characterized by a change of the order parameter from zero to a nonzero value. Since we are considering our system in the vicinity of a phase transition, $\left| \Psi \right|$ has a small value and hence we treat the hopping term in (\ref{H_one}) as a perturbation. Thus, (\ref{H_one}) decomposes according to $\hat{H}=\hat{H}_0+\hat{V}$ into an unperturbed Hamiltonian
\begin{equation}\label{H0}
		\hat{H}_0=\frac{U}{2}\left(\hat{n}^{2}-\hat{n}\right) -\mu\hat{n} +tz\Psi^{*}\Psi,
\end{equation}
and the perturbation
\begin{equation}
		\hat{V}=-tz\left(\Psi^{*}\hat{a}+\Psi\hat{a}^{\dagger}\right). \label{perturbation}
\end{equation}

The unperturbed eigenenergies are
\begin{equation}
\label{En}
E_n = \frac{U}{2}\left(n^{2}-n\right)-\mu n + t z |\Psi|^2 ,
\end{equation}
where the quantum number $n=0,1,2,\cdots$ corresponds to the number of bosons per site.

At this point we are interested in evaluating how the perturbation changes the free energy of the system. For this purpose, we must work out the partition function, 
\begin{equation}
\label{Z1}
\mathcal{Z} = \mathrm{Tr}\left[\mathrm{e}^{-\beta \hat{H}}\right],
\end{equation}
in order to obtain the free energy of the system. The quantum-mechanical evolution operator with imaginary time, i.e., $\hat{U}=\mathrm{e}^{-\beta \hat{H}}$, can be factorized according to
\begin{equation}
\hat{U}=\mathrm{e}^{-\beta \hat{H}_0}\hat{U}_{\rm{I}}(\beta),
\end{equation}
where $\hat{U}_{\rm{I}}(\beta)$ is the interaction picture imaginary-time evolution operator. The equation for the time evolution operator in the interaction picture is \cite{sakurai}
\begin{equation}
\label{eq21a}
\frac{d \hat{U}_{\rm{I}} (\tau)}{d \tau} = - \hat{V}_{\rm{I}} (\tau) \hat{U}_{\rm{I}} (\tau),
\end{equation}
with
\begin{equation}
\label{pertua}
\hat{V}_{\rm{I}}(\tau)=\mathrm{e}^{\tau\hat{H}_{0}}\hat{V}\mathrm{e}^{-\tau\hat{H}_{0}}
\end{equation}
and $\hbar=1$.

Equation (\ref{eq21a}) has to be solved with the initial value $\hat{U}_{\rm{I}}(0)=1$. This can be done iteratively, thus allowing the construction of a perturbative expansion. Up to fourth order in the interaction we have \cite{ednilson}
\begin{align}
\label{UI}
	\hat{U}_{\rm{I}}(\beta) &\approx& \hat{\mathbb{I}} -\int_{0}^{\beta} d\tau_1 \hat{V}_{\rm{I}} (\tau_1) + \int_{0}^{\beta} d\tau_1 \int_{0}^{\tau_1} d\tau_2 \hat{V}_{\rm{I}} (\tau_1) \hat{V}_{\rm{I}} (\tau_2) -\int_{0}^{\beta} d\tau_1 \int_{0}^{\tau_1} d\tau_2 \int_{0}^{\tau_2} d\tau_3 \hat{V}_{\rm{I}} (\tau_1) \hat{V}_{\rm{I}} (\tau_2) \hat{V}_{\rm{I}} (\tau_3) \nonumber \\
&&+\int_{0}^{\beta} d\tau_1 \int_{0}^{\tau_1} d\tau_2 \int_{0}^{\tau_2} d\tau_3 \int_{0}^{\tau_3} d\tau_4 \hat{V}_{\rm{I}} (\tau_1) \hat{V}_{\rm{I}} (\tau_2) \hat{V}_{\rm{I}} (\tau_3) \hat{V}_{\rm{I}} (\tau_4).
\end{align}

It turns out that for the perturbative Hamiltonian in (\ref{perturbation}) all odd-order terms in (\ref{UI}) vanish. Therefore, we can restrict ourselves to the calculation of the zeroth, second, and fourth-order terms in (\ref{UI}).

Making use of the time-evolution operator in the interaction picture $\mathcal{Z} = \Tr{\expo{-\beta \hat{H}_0}\hat{U}_{\rm{I}}(\beta)}$, we calculate the partition function,
\begin{equation}
 Z=\sum_{n=0}^{\infty} \expo{-\beta E_n}\bra{n} \hat{U}_{\rm{I}}(\beta) \ket{n},
\end{equation}
with the single-site eigenstates $\ket{n}$ corresponding to the eigenvalues in Eq. (\ref{En}). The zeroth-order term is
\begin{equation}
\mathcal{Z}^{(0)} = \sum_{n=0}^{\infty}\mathrm{e}^{-\beta E_{n}}.
\end{equation}
The second and fourth-order terms $\mathcal{Z}^{(2)}$ and $\mathcal{Z}^{(4)}$ are calculated in detail in App. \ref{appendix1}.

From the partition function, we then evaluate the free energy,
\begin{equation}\label{F}
    \mathcal{F} = -\frac{1}{\beta}\ln \mathcal{Z} .
\end{equation}
Up to fourth order we get
\begin{equation}\label{F4th}
\mathcal{F}\approx-\frac{1}{\beta} \left[\ln \mathcal{Z}^{(0)} + \frac{\mathcal{Z}^{(2)}}{\mathcal{Z}^{(0)}} + \frac{\mathcal{Z}^{(4)}}{\mathcal{Z}^{(0)}} -\frac{1}{2} \left( \frac{\mathcal{Z}^{(2)}}{\mathcal{Z}^{(0)}}\right)^2 \right].
\end{equation}

Therefore, by comparing (\ref{landauexpansion}) and (\ref{F4th}) we read off the Landau expansion coefficients
\begin{subequations}
	\begin{align}
		a_0&=-\frac{1}{\beta}\ln \mathcal{Z}^{(0)} , \label{a01}
		\\
		a_2&=-\frac{1}{\beta} \frac{1}{|\Psi|^2} \frac{\mathcal{Z}^{(2)}}{\mathcal{Z}^{(0)}}, \label{a21}
		\\
		a_{4}&=-\frac{1}{\beta} \frac{1}{|\Psi|^4} \left[  \frac{\mathcal{Z}^{(4)}}{\mathcal{Z}^{(0)}} -\frac{1}{2} \left( \frac{\mathcal{Z}^{(2)}}{\mathcal{Z}^{(0)}}\right)^2 \right]. \label{a41}
	\end{align}
\end{subequations}

At zero temperature, we obtain results which are equivalent to RSPT. In particular, the Landau expansion coefficients reduce to
\begin{subequations}
	\begin{gather}
	a_0=E_n - t z |\Psi|^2, \label{a0}
	\\
	a_2=t z + (t z)^2 \left( \frac{n+1}{\Delta_{n,n+1}} + \frac{n}{\Delta_{n,n-1}}\right), \label{a2}
	\\
	a_{4}=\left(tz\right)^{4}\left[\frac{n\left(n-1\right)}{\Delta_{n,n-1}^{2}\Delta_{n,n-2}}+\frac{\left(n+1\right)\left(n+2\right)}{\Delta_{n+1,n}^{2}\Delta_{n,n+2}}+\frac{n{}^{2}}{\Delta_{n-1,n}^{3}}+\frac{\left(n+1\right)^{2}}{\Delta_{n+1,n}^{3}}+\frac{n\left(n+1\right)}{\Delta_{n+1,n}^{2}\Delta_{n-1,n}}+\frac{n\left(n+1\right)}{\Delta_{n,n-1}^{2}\Delta_{n+1,n}}\right], \label{a4}
	\end{gather}
\end{subequations}
where $\Delta_{i,j}\equiv E_i-E_j$.

\begin{figure}[t]
	\centering
	\includegraphics[width=.6\columnwidth]{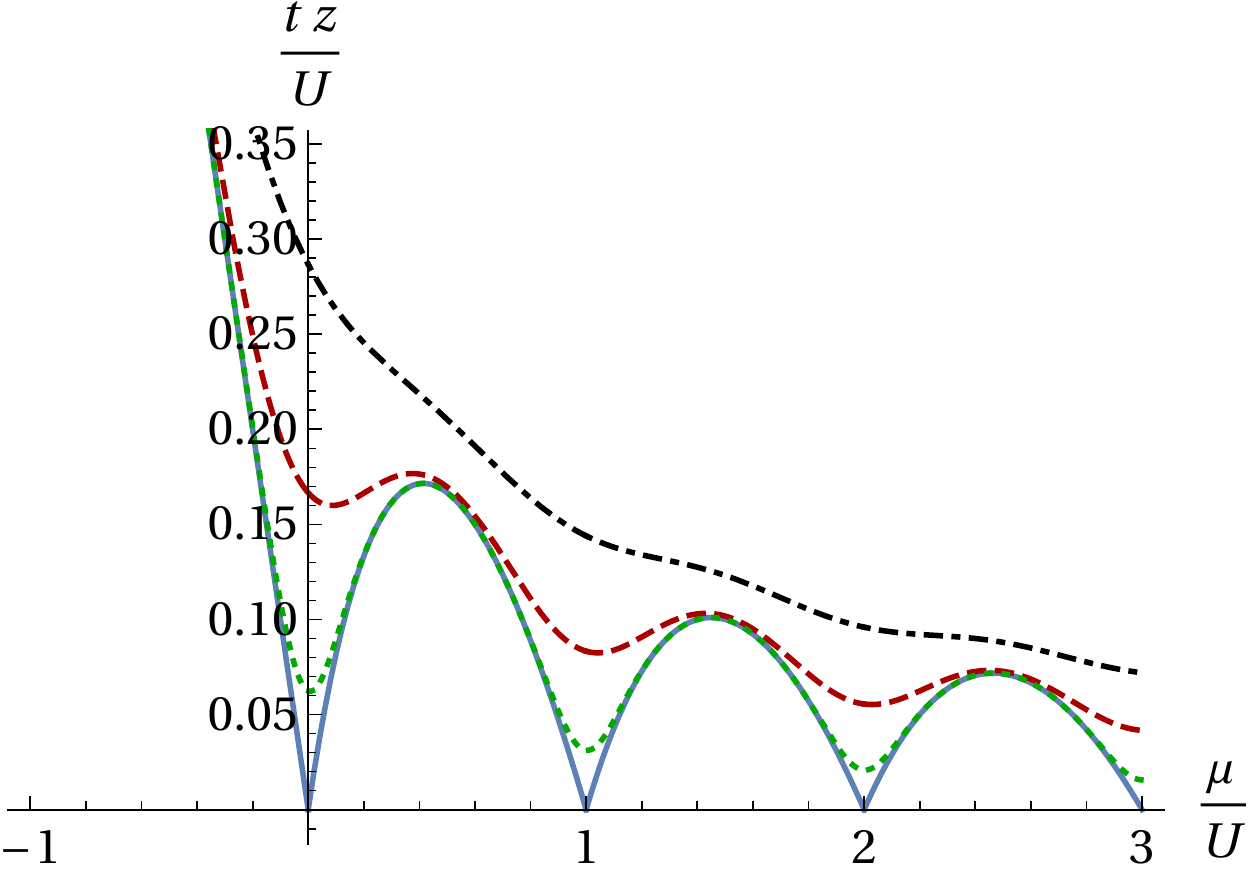}
	\caption{(Color online) Phase diagram for the inverse temperatures $\beta=5/U$ (dotted-dashed black), $\beta=10/U$ (dashed red), $\beta=30/U$ (dotted green), and $\beta \rightarrow \infty$ (continuous blue).}
	\label{pb_manyT}
\end{figure}

The explicit solution for $a_2=0$ gives the phase boundaries in Fig. \ref{pb_manyT}, as in \cite{ednilson}. From Fig. \ref{pb_manyT} we read off that thermal fluctuations mainly affect the phase boundary between two Mott lobes due to fluctuations in the number of bosons per site.

\subsection{\label{inconsistencies}Nondegenerate perturbation theory inconsistency}
As already pointed out, NDPT is expected to exhibit degeneracy-related problems. Indeed, by directly observing the coefficient denominators in (\ref{a2}) and (\ref{a4}) we clearly identify such degeneracy problems. Whenever $\mu/U$ becomes an integer $n$, there is an equality between two consecutive energy values, for instance $E_n$ and $E_{n+1}$, thus characterizing a divergence in these expressions.

According to (\ref{landauexpansion}), we can consider the Landau expansion up to fourth order for the free energy in the vicinity of a phase transition. Extremizing (\ref{landauexpansion}) with respect to the order parameter leads to
\begin{equation}
\frac{\partial \mathcal{F}}{\partial |\Psi|^2} = a_2 + 2 a_4 |\Psi|^2 = 0,
\end{equation}
with the solution in the superfluid phase
\begin{equation}
\label{ordpara}
 |\Psi|^2 = -\frac{a_2}{2 a_4}.
\end{equation}

\begin{figure}[t]
	\centering
	\includegraphics[width=.6\columnwidth]{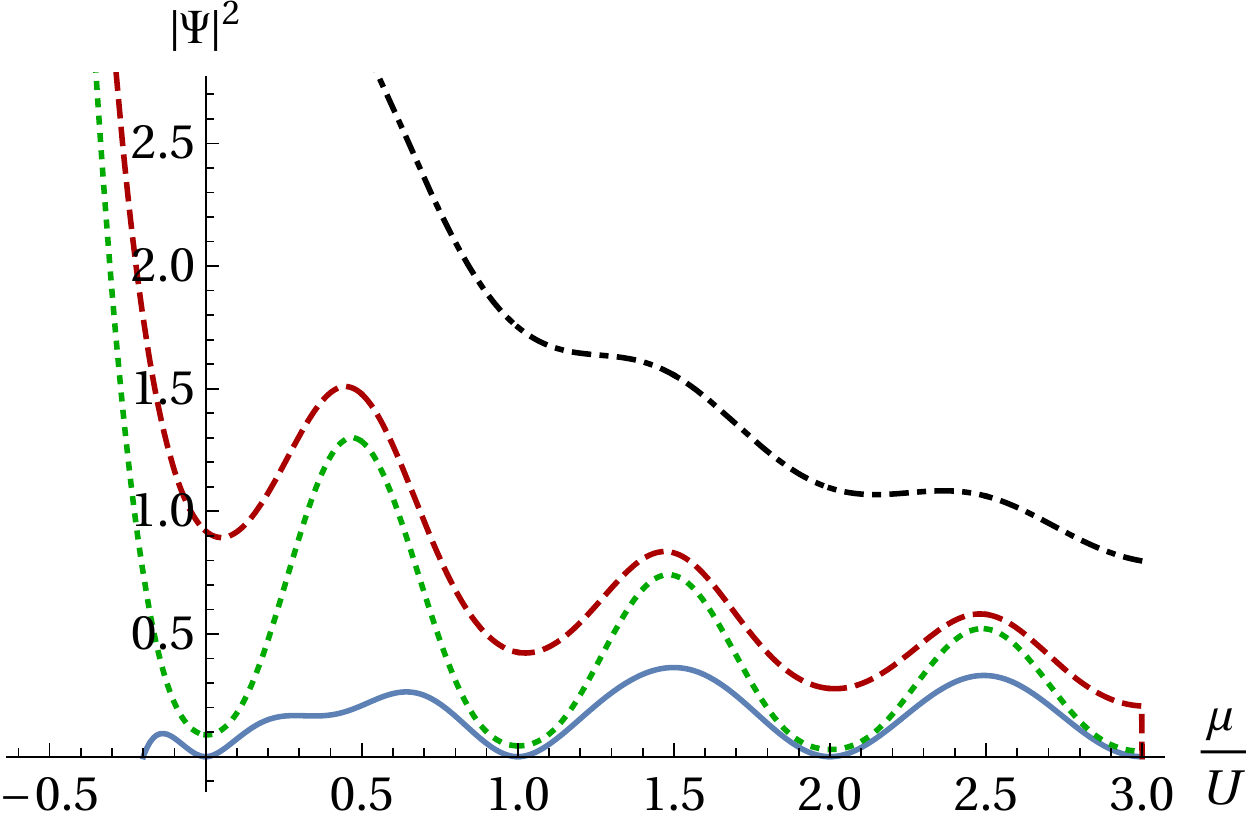}
	\caption{(Color online) Order parameter via NDPT from (\ref{ordpara}) as a function of $\mu/U$ for $tz/U = 0.2$ as well as $\beta=5/U$ (dotted-dashed black), $\beta=10/U$ (dashed red), $\beta=30/U$ (dotted green), and $\beta \rightarrow \infty$ (continuous blue).}
	\label{order_parameter}
\end{figure}

The plot of $|\Psi|^2$ as a function of $\mu/U$ making use of (\ref{a21}) and (\ref{a41}) is interesting for our purposes since it reveals an unphysical behavior, which is a consequence of NDPT: the order parameter approaches zero at a point where no phase transition occurs. Fig. \ref{order_parameter} shows equation (\ref{ordpara}) for $tz/U=0.2$ for a varying chemical potential. We observe that, indeed, the OP is well-behaved in most parts of the diagram. However, it also shows an inconsistency: at integer values of $\mu/U$ the order parameter for the zero-temperature system goes to zero, while for $T>0$ it mimics the zero-temperature behavior by decreasing its values despite of not vanishing.

Since, for finite temperatures, NDPT also shows a similar unphysical behavior typical of RSPT, in the following section we demonstrate how such problems can be fixed at finite temperature.

\section{\label{sec3}Degenerate Approach}
In this section, we introduce our method which consists of a degenerate perturbative calculation making use of projection operators. As we aim at describing the region between two neighboring Mott lobes, we define a subspace of the Hilbert space which is composed of two degenerate states with $n$ and $n+1$ particles via the projection operator
\begin{equation} 
    \hat{P} = \vert n \rangle \langle n \vert + \vert n+1 \rangle \langle n+1 \vert,
\end{equation}
this way the corresponding complementary operator is given by 
\begin{equation} 
    \hat{Q} = \sum_{m \notin P} \vert m \rangle \langle m \vert.
\end{equation}

We begin our analysis by considering the one-site mean-field Hamiltonian (\ref{H_one}) and regard, as in Sec. \ref{PT}, the hopping term (\ref{perturbation}) as a perturbation for (\ref{H0}). We multiply both sides of the perturbation by the identity operator, $\hat{\mathbb{I}} = \hat{P}+\hat{Q}$, and get
\begin{equation}
\hat{H}=\hat{H}_{0}+\left(\hat{P}+\hat{Q}\right)\hat{V}\left(\hat{P}+\hat{Q}\right)
\end{equation}

Then we define the new unperturbed Hamiltonian and the new perturbation as
\begin{subequations}
\begin{align}
\hat{\mathcal{H}}_{0}&\equiv\hat{H}_{0}+\hat{P}\hat{V}\hat{P}, \label{newH0}\\
\hat{\mathcal{V}}&\equiv\hat{P}\hat{V}\hat{Q}+\hat{Q}\hat{V}\hat{P}+\hat{Q}\hat{V}\hat{Q}. \label{newV}
\end{align}
\end{subequations}

The Hamiltonian in equation (\ref{newH0}), written in the basis of the unperturbed eigenstates, is a block diagonal matrix, whose only nondiagonal block is
\begin{equation}
\hat{\mathcal{H}}_{0}^{\rm(nd)}=\begin{pmatrix} E_{n} & -t z \Psi \sqrt{n+1} \\
 -t z \Psi^{*}\sqrt{n+1} & E_{n+1} \end{pmatrix}.
\end{equation}
Its eigenvalues and eigenstates are given by
\begin{subequations}
\begin{align}
\mathcal{E}_{\pm} &= \frac{E_{n}+E_{n+1}}{2}
\pm\frac{1}{2}\sqrt{\left(E_{n}-E_{n+1}\right)^{2}+4 t^2 z^2 |\Psi|^2\left(n+1\right)}, \\
	\vert\Phi_{\pm}\rangle&=\left(1+\frac{\left|\mathcal{E}_{\pm}-E_{n}\right|^{2}}{t^2 z^2 |\Psi|^2 \left(n+1\right)}\right)^{-1/2}\left(\vert n\rangle+\frac{E_{n}-\mathcal{E}_{\pm}}{t z\sqrt{|\Psi|^{2}\left(n+1\right)}}\vert n+1\rangle\right). \label{PHI+-}
\end{align}
\end{subequations}

As pointed out in Sec. \ref{PT}, we must evaluate the partition function (\ref{Z1}) in order to calculate the free energy (\ref{F}). The only difference is that now we are working with the new unperturbed Hamiltonian (\ref{newH0}) and the new perturbation (\ref{newV}). With this the time evolution operator now reads
\begin{equation}
\hat{U}=\mathrm{e}^{-\beta \hat{\mathcal{H}}_0}\hat{\mathcal{U}}_{\rm{I}}.
\end{equation}
The equation for the imaginary-time-evolution operator in the interaction picture is
\begin{equation}
\label{eq21}
\frac{d \hat{\mathcal{U}}_{\rm{I}} (\tau)}{d \tau} = - \hat{\mathcal{V}}_{\rm{I}} (\tau) \hat{\mathcal{U}}_{\rm{I}} (\tau),
\end{equation}
with
\begin{equation}
\label{pertu}
\hat{\mathcal{V}}_{\rm{I}}(\tau)=\mathrm{e}^{\tau\hat{\mathcal{H}}_{0}}\left(\hat{P}\hat{V}\hat{Q}+\hat{Q}\hat{V}\hat{P}+\hat{Q}\hat{V}\hat{Q}\right)\mathrm{e}^{-\tau\hat{\mathcal{H}}_{0}}.
\end{equation}

The solution for equation (\ref{eq21}) with the initial condition $\hat{\mathcal{U}}_{\rm{I}}(0)=1$ up to second order is given by
\begin{equation}\label{U}
	\hat{\mathcal{U}}_{\rm{I}}(\beta) = \hat{\mathbb{I}}-\int_{0}^{\beta} d\tau_1 \hat{\mathcal{V}}_{\rm{I}} (\tau_1) + \int_{0}^{\beta} d\tau_1 \int_{0}^{\tau_1} d\tau_2 \hat{\mathcal{V}}_{\rm{I}} (\tau_1) \hat{\mathcal{V}}_{\rm{I}} (\tau_2).
\end{equation}

Evaluating the partition function $\mathcal{Z}=\Tr{\expo{-\beta
\hat{\mathcal{H}}_0}\hat{\mathcal{U}}_{\rm{I}}(\beta)}$, we have \begin{align}
	\label{Z} \mathcal{Z}&= \expo{-\beta \mathcal{E}_+}\bra{\Phi_+}
	\hat{\mathcal{U}}_{\rm{I}}(\beta) \ket{\Phi_+} + \expo{-\beta
	\mathcal{E}_-}\bra{\Phi_-} \hat{\mathcal{U}}_{\rm{I}}(\beta)
	\ket{\Phi_-} +\sum_{m \in Q} \expo{-\beta E_m}\bra{m}
	\hat{\mathcal{U}}_{\rm{I}}(\beta) \ket{m}.  \end{align} The
	zeroth-order term in (\ref{U}) yields in (\ref{Z}) \begin{equation}
		\label{Z0} \mathcal{Z}^{(0)} = \expo{-\beta \mathcal{E}_+} +
		\expo{-\beta \mathcal{E}_-} + \sum_{m \in Q} \expo{-\beta E_m}.
	\end{equation} Furthermore, we read off from (\ref{perturbation}),
	(\ref{pertu}) and (\ref{newV}) that the first-order contribution in
	(\ref{Z}) must vanish.

Finally, the second-order term, which is calculated in detail in Appendix
\ref{appendix}, gives \begin{align} \label{Z2}
	\mathcal{Z}^{(2)}&=t^{2}z^{2}\vert\Psi\vert^{2}\Bigg\{
		(n+2)\beta\Bigg[\big|\braket{\Phi_+}{n+1}\big|^{2}\Bigg(\frac{\mathrm{e}^{-\beta\mathcal{E}_{+}}-\mathrm{e}^{-\beta
		E_{n+2}}}{\Delta_{n+2,+}}\Bigg)+\big|\braket{\Phi_-}{n+1}\big|^{2}\Bigg(\frac{\mathrm{e}^{-\beta\mathcal{E}_{-}}-\mathrm{e}^{-\beta
		E_{n+2}}}{\Delta_{n+2,-}}\Bigg)\Bigg] \nonumber \\
		&+n\beta\Bigg[\big|\braket{\Phi_+}{n}\big|^{2}\Bigg(\frac{\mathrm{e}^{-\beta\mathcal{E}_{+}}-\mathrm{e}^{-\beta
		E_{n-1}}}{\Delta_{n-1,+}}\Bigg)+\big|\braket{\Phi_-}{n}\big|^{2}\Bigg(\frac{\mathrm{e}^{-\beta\mathcal{E}_{-}}-\mathrm{e}^{-\beta
		E_{n-1}}}{\Delta_{n-1,-}}\Bigg)\Bigg] \nonumber \\
		&+\sum_{\substack{m \in Q \\ m \neq
		n-1}}(m+1)\Bigg(\frac{\mathrm{e}^{-\beta
		E_{m+1}}-\mathrm{e}^{-\beta
		E_{m}}}{\Delta_{m,m+1}^{2}}-\frac{\beta\mathrm{e}^{-\beta
		E_{m}}}{\Delta_{m,m+1}}\Bigg)+\sum_{\substack{m \in Q \\ m\neq
		n+2}}m\Bigg(\frac{\mathrm{e}^{-\beta
		E_{m-1}}-\mathrm{e}^{-\beta
		E_{m}}}{\Delta_{m,m-1}^{2}}-\frac{\beta\mathrm{e}^{-\beta
		E_{m}}}{\Delta_{m,m-1}}\Bigg)\Bigg\}, \end{align} where we have
		introduced the abbreviation $\Delta_{i,\pm} \equiv E_i -
		\mathcal{E}_\pm$.

From Eq. (\ref{Z2}), we observe that the difference between the degenerate
energies $E_n$ and $E_{n+1}$ will no longer appear in the denominator of the
free energy as it did in the NDPT treatment, thus solving the
degeneracy-related problems discussed above.

\subsection{\label{CD}Condensate Density}
\begin{figure}[H] \centering
		
		%\begin{subfigure}{.45\columnwidth}
		%	\includegraphics[width=\columnwidth]{condensate_density_n=0_finiteT5.pdf}
		%	\caption{}
		%\label{n=0cd_finiteT5} \end{subfigure} \qquad
		\begin{subfigure}{.45\columnwidth}
			\includegraphics[width=\columnwidth]{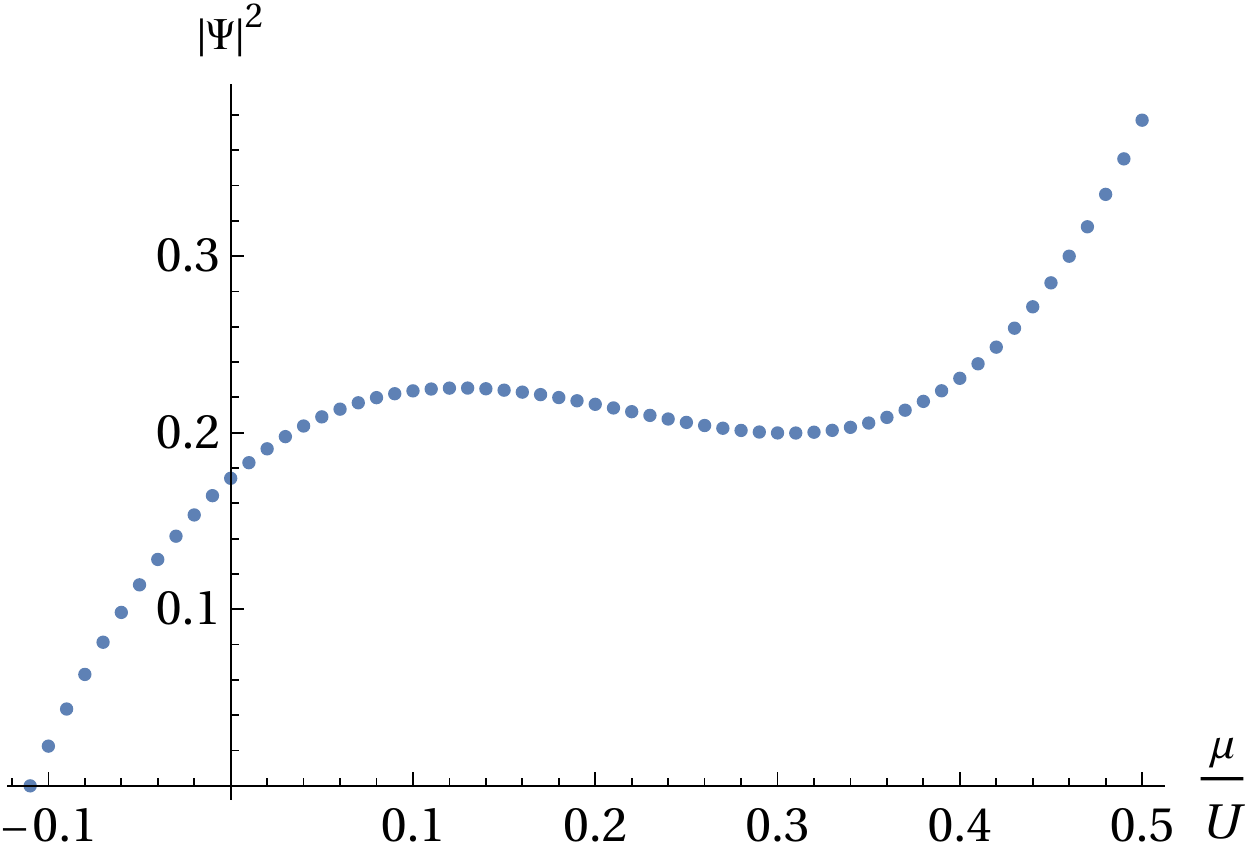}
			\caption{}
		\label{n=0cd_finiteT10} \end{subfigure} \qquad
		\begin{subfigure}{.45\columnwidth}
			\includegraphics[width=\columnwidth]{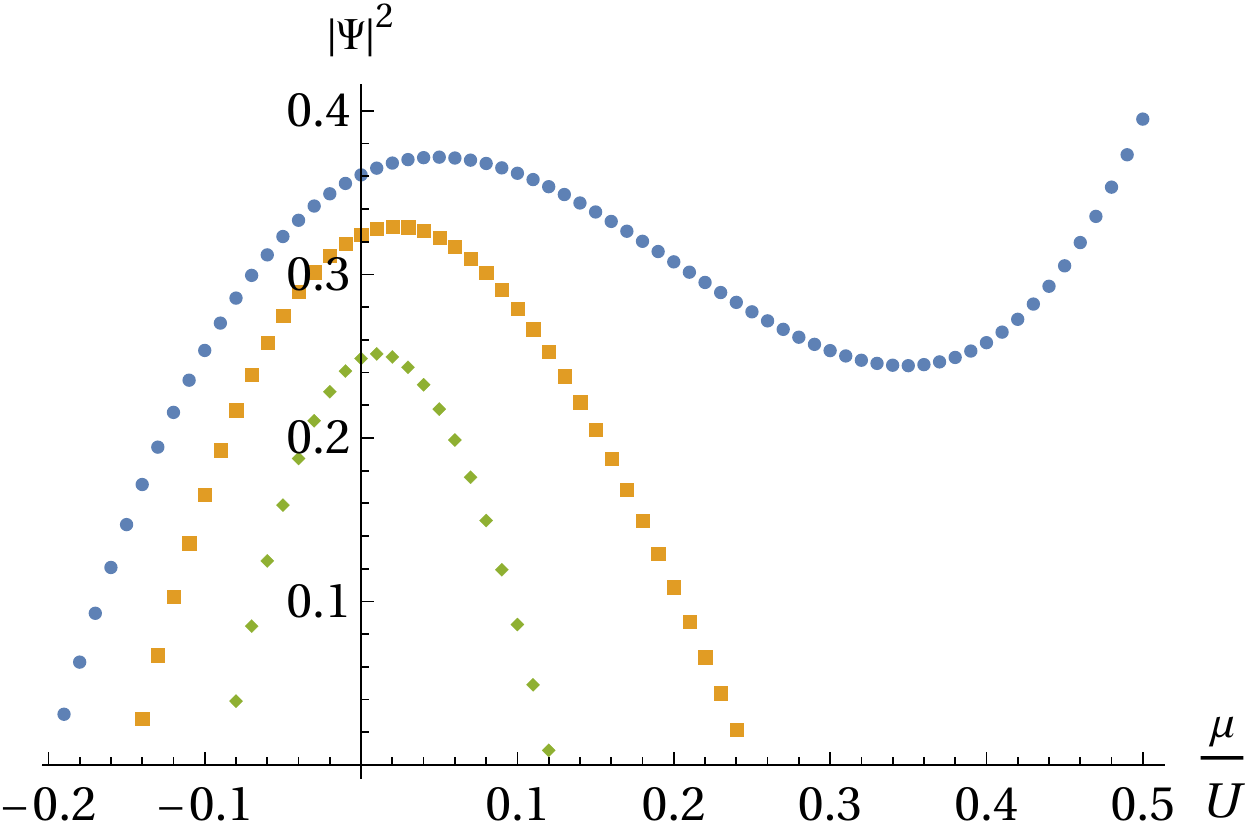}
			\caption{}
		\label{n=0cd_finiteT30} \end{subfigure} \qquad
		\begin{subfigure}{.45\columnwidth}
			\includegraphics[width=\columnwidth]{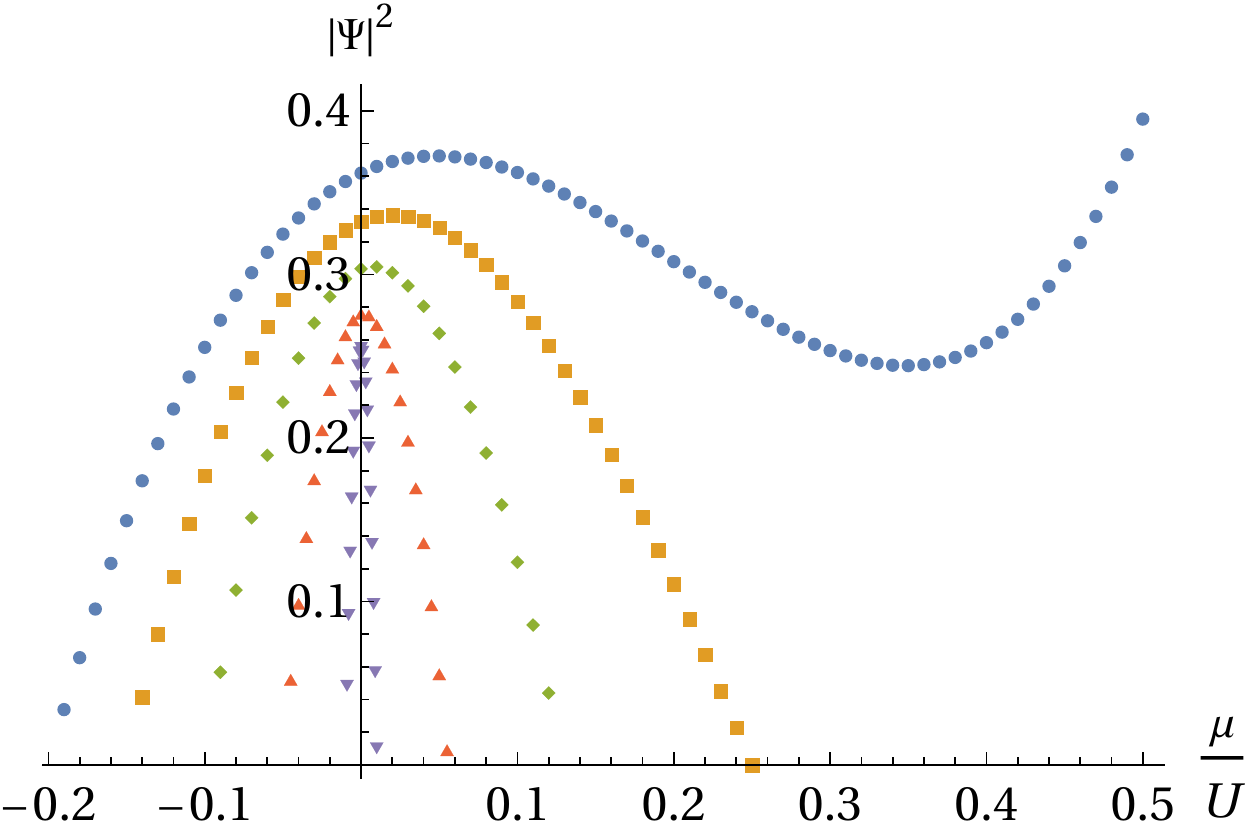}
			\caption{}
		\label{n=0cd_zeroT} \end{subfigure}
		
		\caption{(Color online) Condensate densities near degeneracy evaluated from FTDPT via (\ref{eq31})
between $n=0$ and $1$, for three different temperatures: (a) $\beta=10/U$, (b) $\beta=30/U$, and (c) $T=0$. Different point styles correspond to different hopping: $tz/U=0.2$ (blue circles), $tz/U=0.15$ (orange squares), $tz/U=0.1$ (green rhombuses), $tz/U=0.05$ (red triangles), and $tz/U=0.01$ (purple inverted triangles).} \label{n=0cond_dens} \end{figure}

\begin{figure}[H] \centering
		\begin{subfigure}{.45\columnwidth}
			\includegraphics[width=\columnwidth]{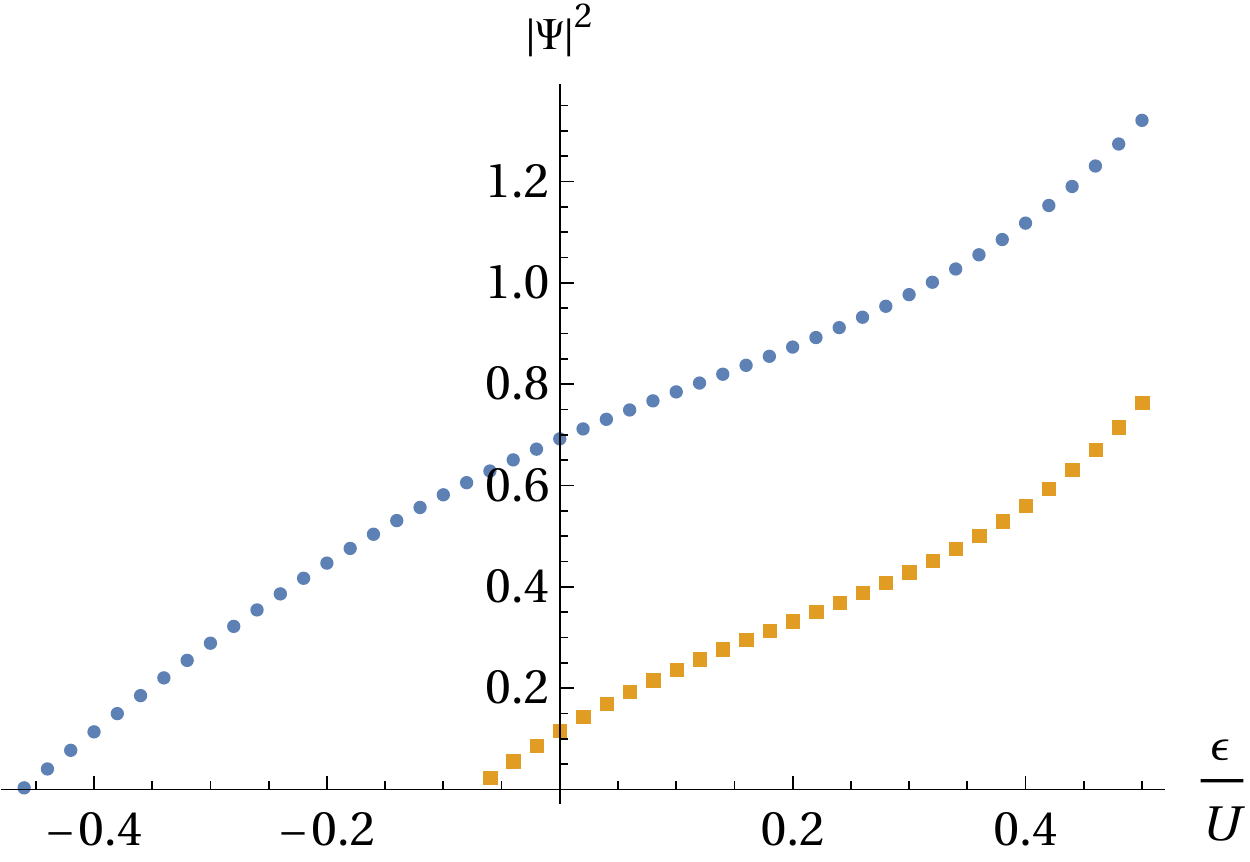}
			\caption{}
		\label{cd_finiteT5} \end{subfigure} \qquad
		\begin{subfigure}{.45\columnwidth}
			\includegraphics[width=\columnwidth]{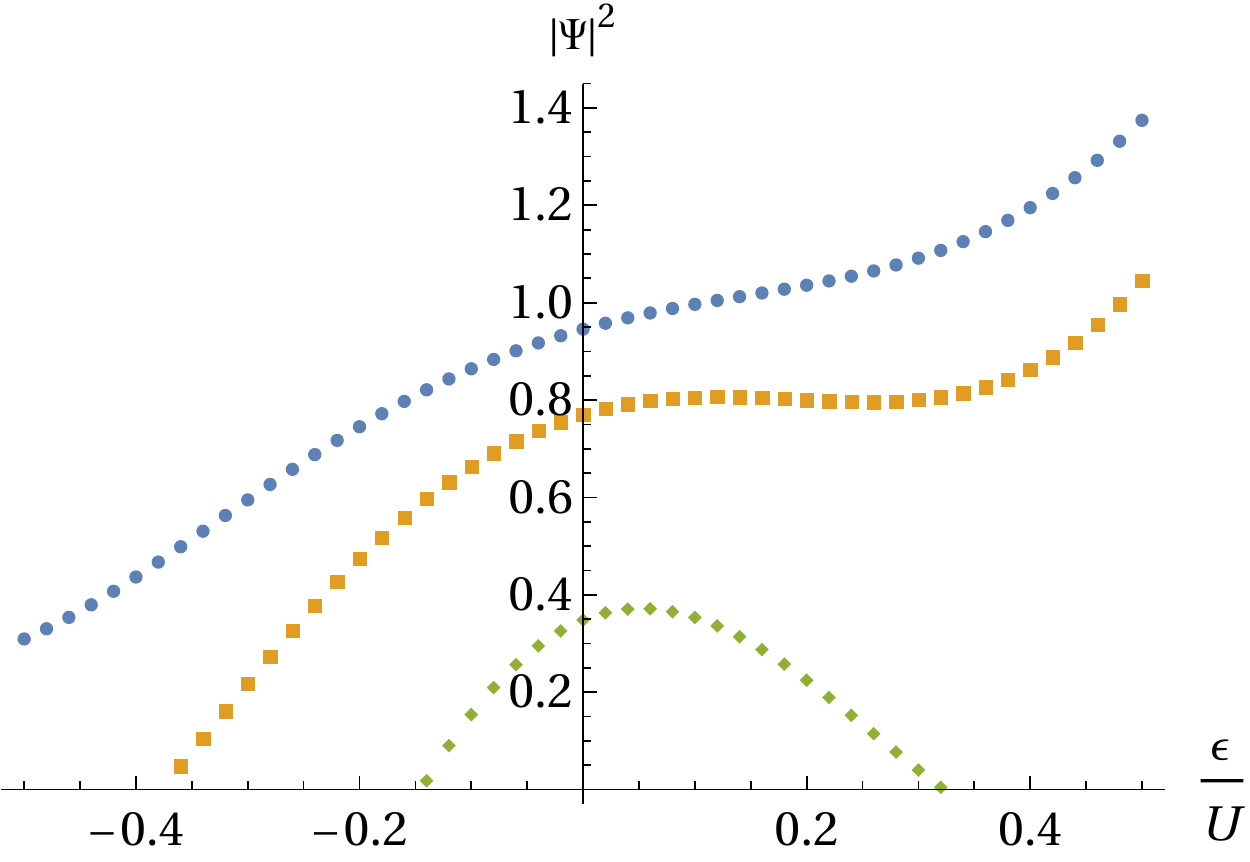}
			\caption{}
		\label{cd_finiteT10} \end{subfigure} \qquad
		\begin{subfigure}{.45\columnwidth}
			\includegraphics[width=\columnwidth]{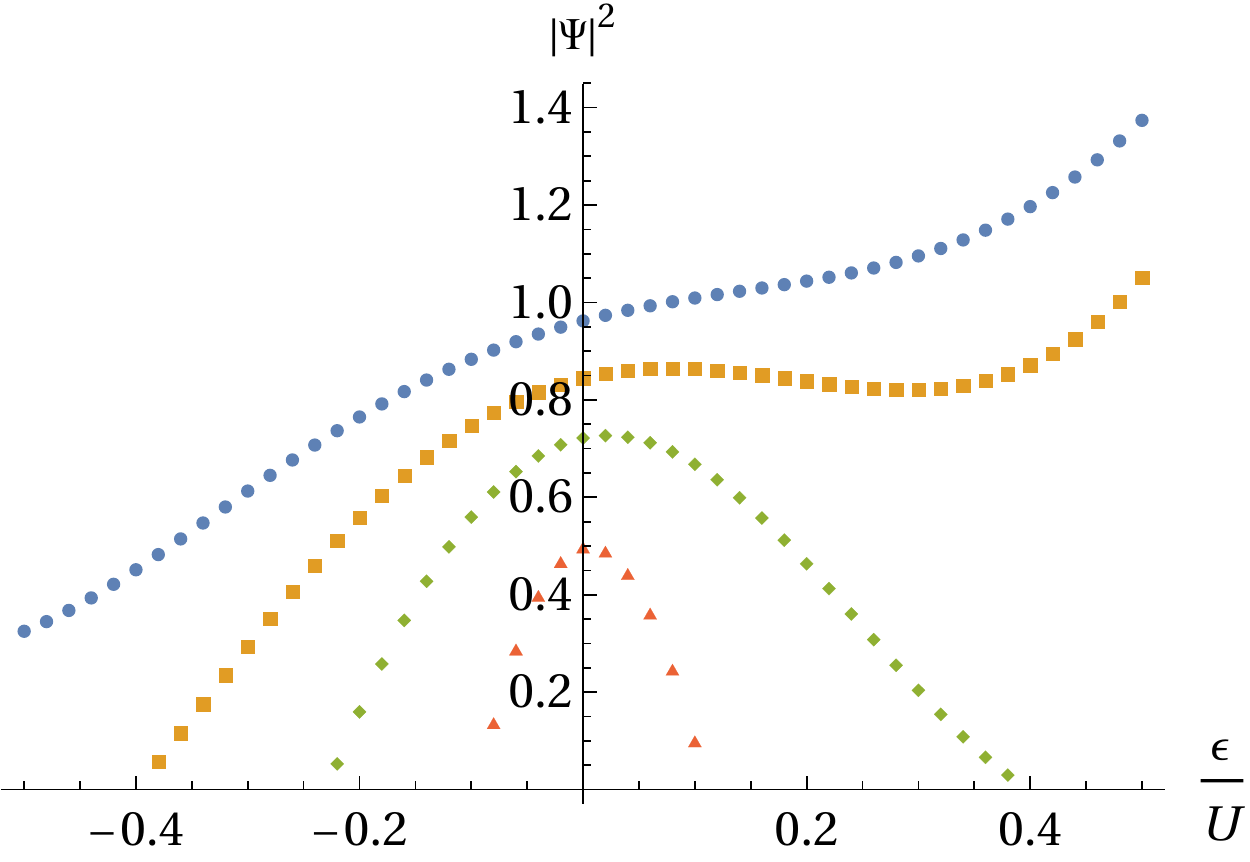}
			\caption{}
		\label{cd_finiteT30} \end{subfigure} \qquad
		\begin{subfigure}{.45\columnwidth}
			\includegraphics[width=\columnwidth]{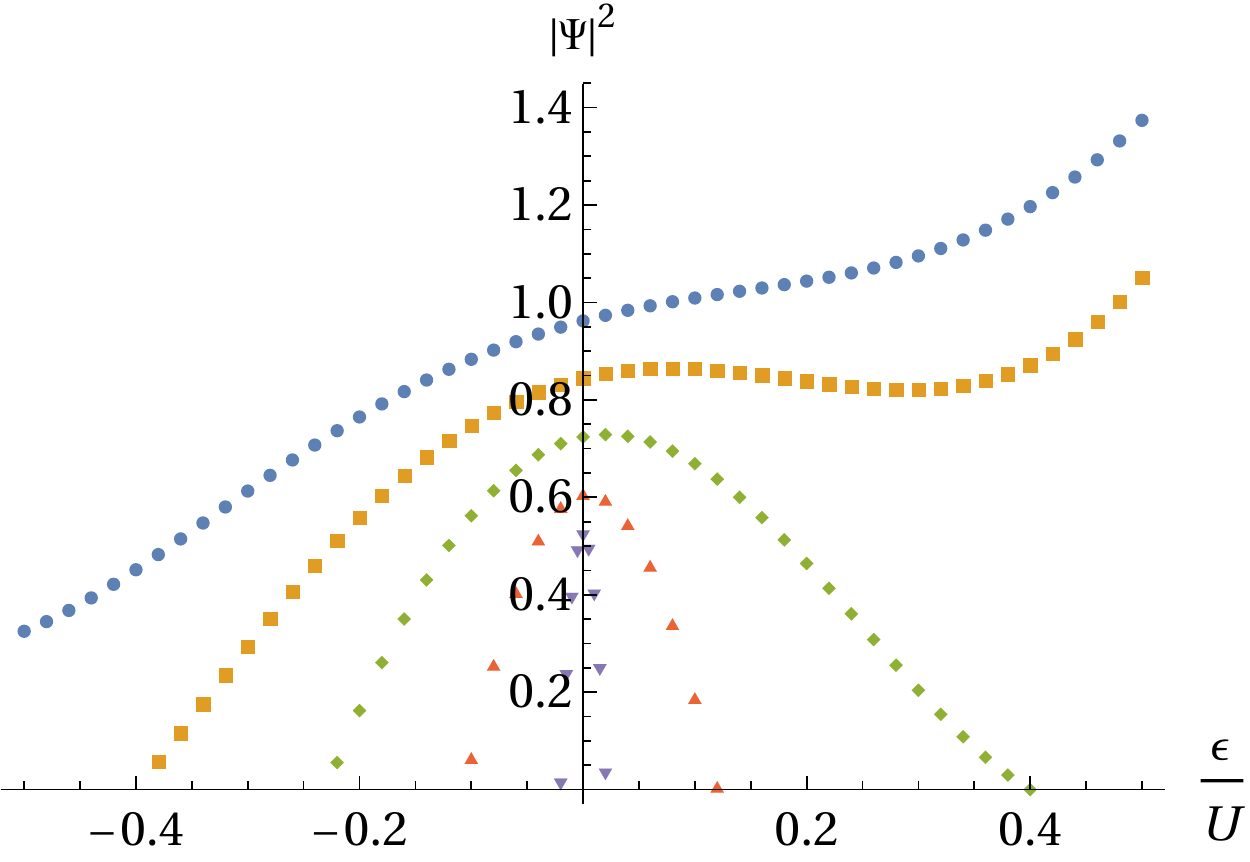}
			\caption{}
		\label{cd_zeroT} \end{subfigure}
		\caption{(Color online) Condensate densities near degeneracy evaluated from FTDPT via (\ref{eq31})
		between the first and second Mott lobes, $\mu = U+\epsilon$, for four different temperatures: (a) $\beta=5/U$, (b) $\beta=10/U$, (c) $\beta=30/U$, and (d) $T=0$. Different point styles correspond to different hopping: $tz/U=0.2$ (blue circles), $tz/U=0.15$ (orange squares), $tz/U=0.1$ (green rhombuses), $tz/U=0.05$ (red triangles), and $tz/U=0.01$ (purple inverted triangles).} \label{cond_dens}
	\end{figure}

Now we turn our attention to the calculation of the condensate density, which
turns out to coincide with the superfluid density in the mean-field
approximation \cite{martin}. Our degenerate approach, up to second order,
results in the partition function given by $\mathcal{Z}= \mathcal{Z}^{(0)} +
\mathcal{Z}^{(2)}$ with (\ref{Z0}) and (\ref{Z2}), which is free from any
divergence despite of the degeneracies. From the partition function, we obtain
for the system free energy (\ref{F}) \begin{equation} \mathcal{F} =
-\frac{1}{\beta} \left[\ln \mathcal{Z}^{(0)} +
\frac{\mathcal{Z}^{(2)}}{\mathcal{Z}^{(0)}} \right].  \end{equation} Hence, we
evaluate the condensate density $|\Psi|^2$ by evaluating
\begin{equation} \label{eq31} \frac{\partial \mathcal{F}}{\partial |\Psi|^2} =
0.  \end{equation} We apply this procedure by considering different
temperatures between the Mott lobes $n=0$ and $1$ in Fig.
\ref{n=0cond_dens}, and between $n=1$ and $2$ in Fig. \ref{cond_dens}.

In order to check the fidelity of the calculated condensate densities we must
observe the phase boundary evaluated by FTDPT, which is given by
\begin{equation} \frac{\partial \mathcal{F}}{\partial
|\Psi|^2}\Bigg|_{\Psi=0}=0.  \end{equation} This procedure leads to the same
phase diagram evaluated by NDPT. From Fig. \ref{pb_manyT}
we read off that for small values of $tz/U$ there are bigger
portions of values of $\mu/U$ where the condensate density can be evaluated,
since we regard the Landau expansion of the order parameter being valid in the
vicinity of the phase transition, i.e. the smaller the hopping, the bigger the
region of the calculated condensate density.  Therefore, we conclude that we
are able to reliably calculate $\left|\Psi\right|^2$ via FTDPT near the phase
boundary in Fig. \ref{n=0cond_dens}. Also, we observe that for $\mu/U=0$
the condensate densities no longer vanish or approach zero as they do when
calculated from NDPT. Regarding the condensate densities calculated between the
first and the second Mott lobes, Fig. \ref{cond_dens}, we also find that the decreasing behavior characteristic of the NDPT in the degeneracy
point $\mu/U=1$ is absent.

\subsection{\label{comparison}Comparison between NDPT and FTDPT}
\begin{figure}[H] \centering \begin{subfigure}{.45\columnwidth}
		\includegraphics[width=\columnwidth]{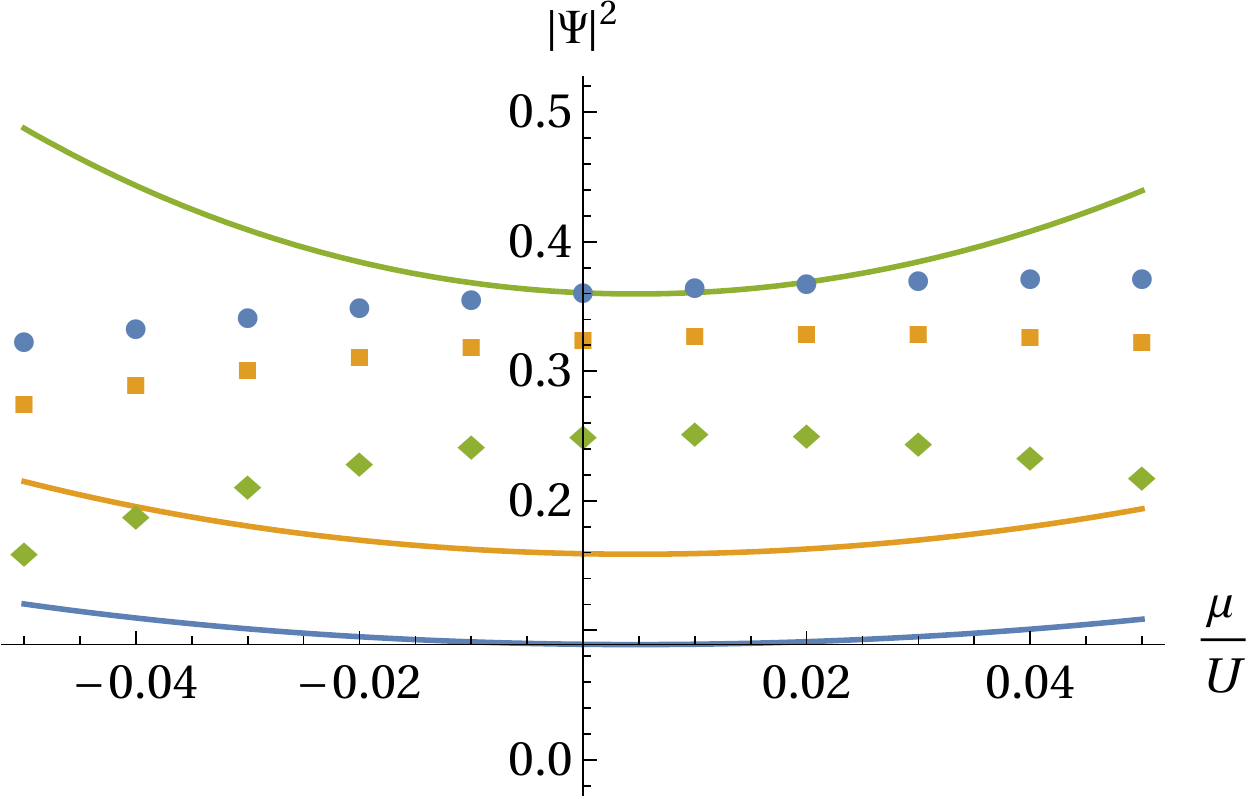} 
			\caption{}
			\label{pl1} \end{subfigure} \qquad
			\begin{subfigure}{.45\columnwidth}
				\includegraphics[width=\columnwidth]{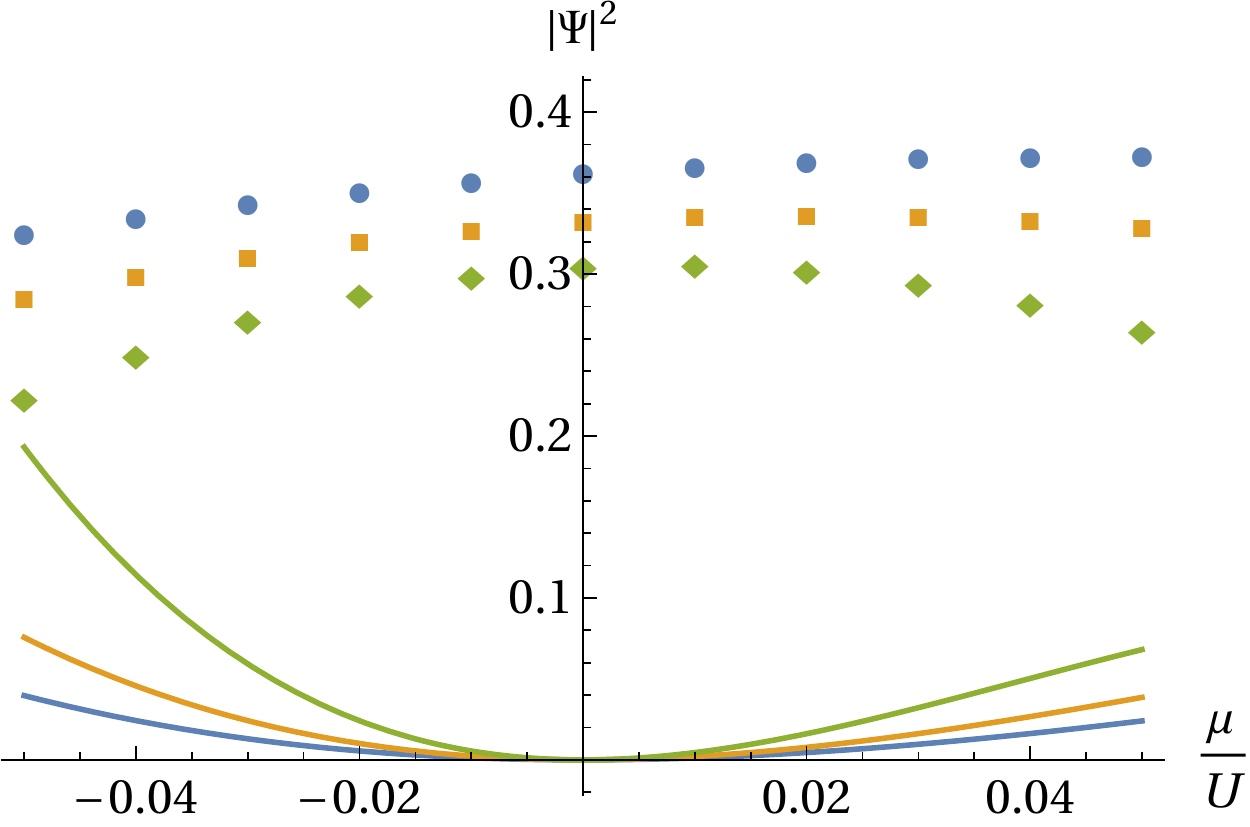}
			\caption{}
			\label{pl2} \end{subfigure}
			\begin{subfigure}{.45\columnwidth}
		\includegraphics[width=\columnwidth]{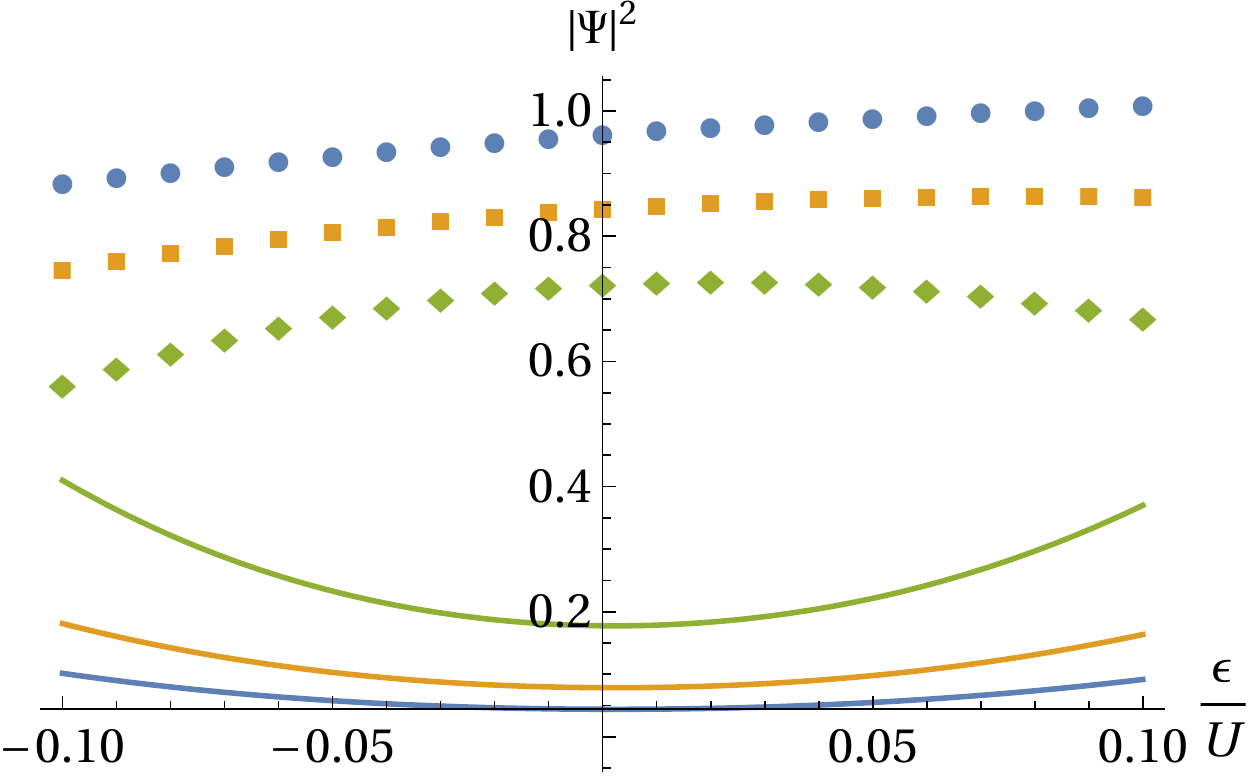} 
			\caption{}
			\label{pl} \end{subfigure} \qquad
			\begin{subfigure}{.45\columnwidth}
				\includegraphics[width=\columnwidth]{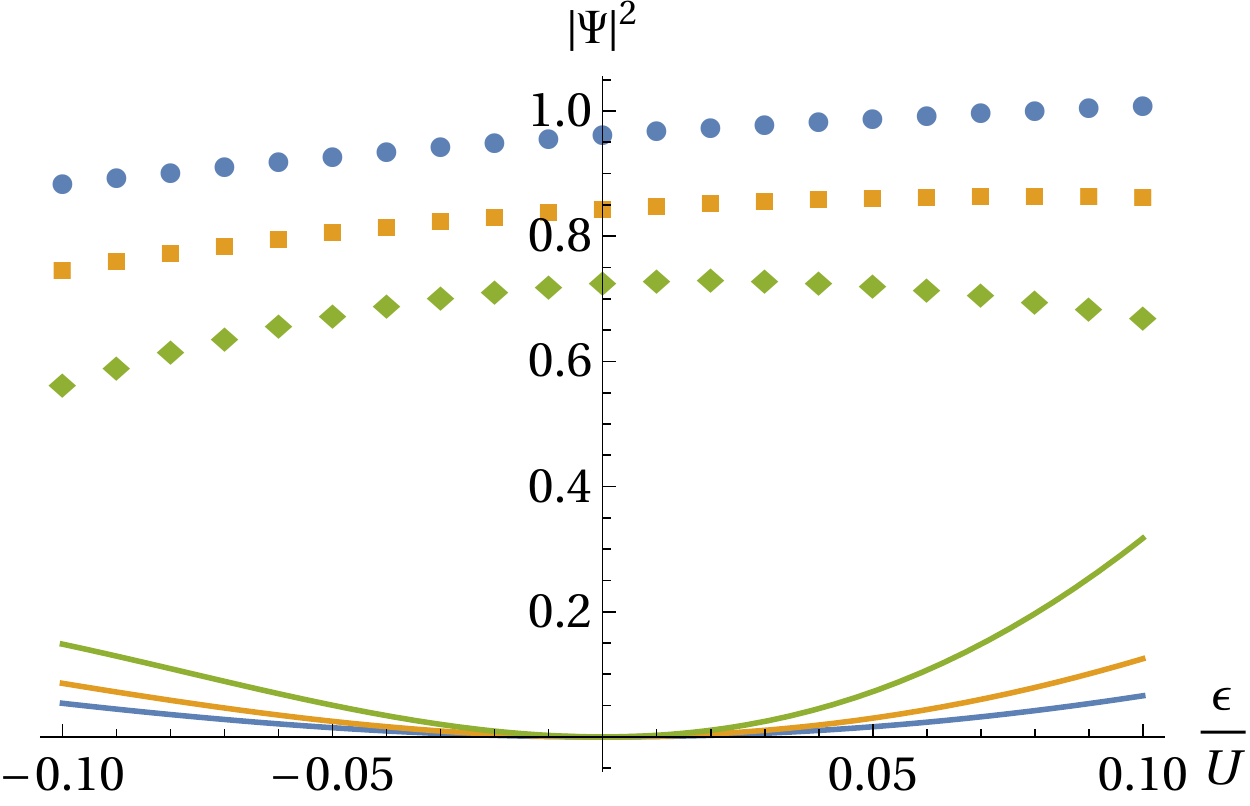}
			\caption{}
			\label{l} \end{subfigure} \caption{(Color online)
			Comparison between the condensate densities calculated
			via FTDPT (points) and NDPT (lines) for the
			temperatures $\beta=30/U$ (left panel) and $T=0$ (right
			panel) for $tz/U=0.2$ (blue circles), $tz/U=0.15$
			(orange squares), and $tz/U=0.1$ (green rhombuses). (a) and 
			(b) correspond to the region between $n=0$ and 1, while (c)
			and (d) correspond to the region between the first and second lobes, with $\mu =
			U+\epsilon$.} \label{d} \end{figure}

Now we turn our attention to the point between two consecutive Mott lobes in
order to analyze the differences between the condensate densities calculated
via NDPT and FTDPT between the Mott lobes $n=0$ and 1, and $n=1$ and 2, as shown in
Fig. \ref{d}. We observe that the NDPT gives
condensate densities that approach zero or have a decreasing behavior at the
degeneracy point, which corresponds to $\mu/U=0$ for the region between $n=0$
and $n=1$ and is depicted in Figs. \ref{pl1} and \ref{pl2}; while for the region between the
first and the second Mott lobes, i.e., $n=1$ and 2, the degeneracy occurs at
$\mu/U=1$ and is depicted in Fig. \ref{d}. Such behavior indicates an
inaccuracy of the theory, since it mimics the unphysical vanishing of the OP
typical of RSPT, which is a direct consequence of not taking into account the
degeneracies that happen in between two consecutive Mott lobes.
While NDPT presents such unphysical behavior due to the incorrect treatment of
degeneracies, FTDPT gives consistent results for the condensate density between
two consecutive Mott lobes. 
	
We observe from Fig. \ref{d} that the condensate densities
calculated via FTDPT, which are represented by the solid lines, do not present
any decreasing behavior in the vicinity of the degeneracy, concluding that they
are consistent in all considered regions of the phase diagram. In particular,
at $\mu/U=0$ and 1 the condensate densities no longer vanish or present a decreasing behavior as they do when
calculated from NDPT. The decreasing behavior presented by the
condensate densities calculated via NDPT can clearly be observed by the solid
lines in Fig. \ref{d}. Such decreasing behavior is a direct
consequence of the incorrect treatment of degeneracies by NDPT, which happens to occur
between two consecutive Mott lobes.

\subsection{\label{density}Particle density}
\begin{figure}[H] \centering
		
		\begin{subfigure}{.45\columnwidth}
			\includegraphics[width=\columnwidth]{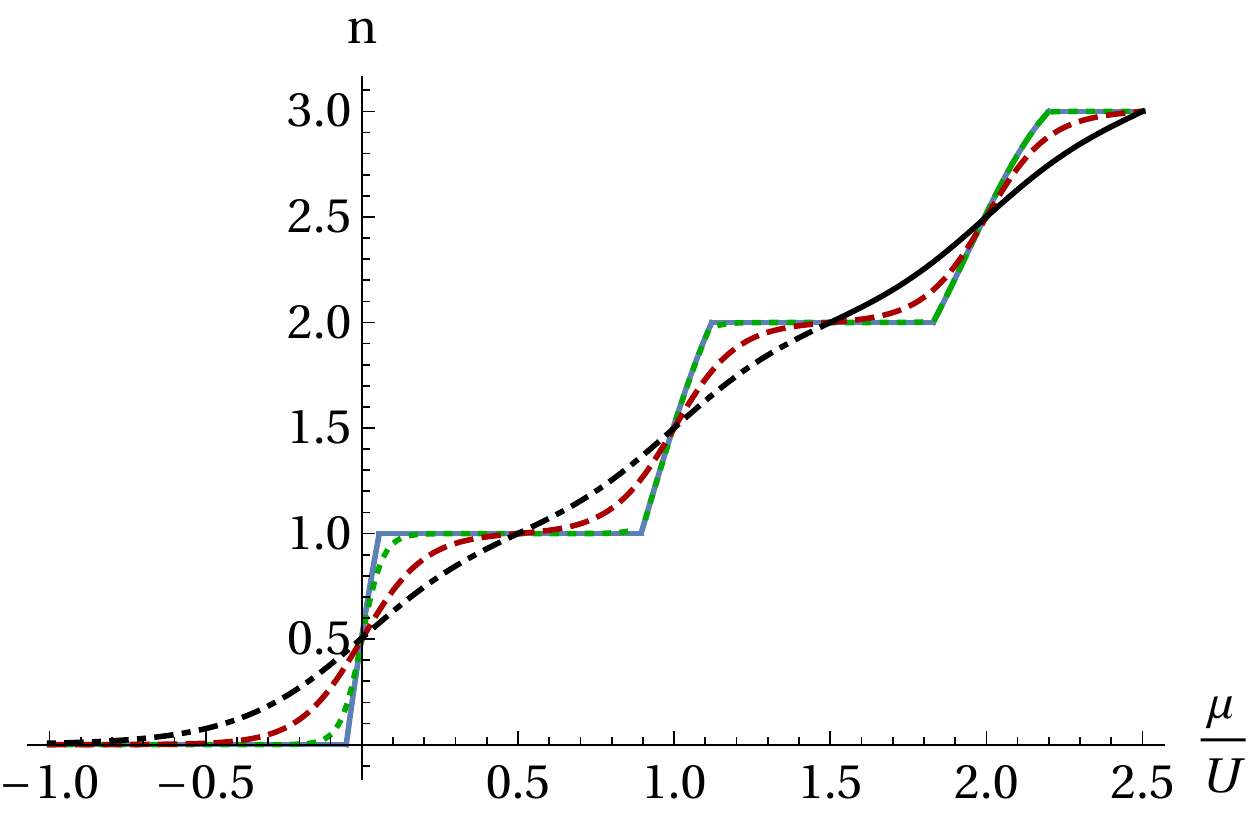} 
			\caption{}
			\label{pd1} \end{subfigure} \qquad
			\begin{subfigure}{.45\columnwidth}
				\includegraphics[width=\columnwidth]{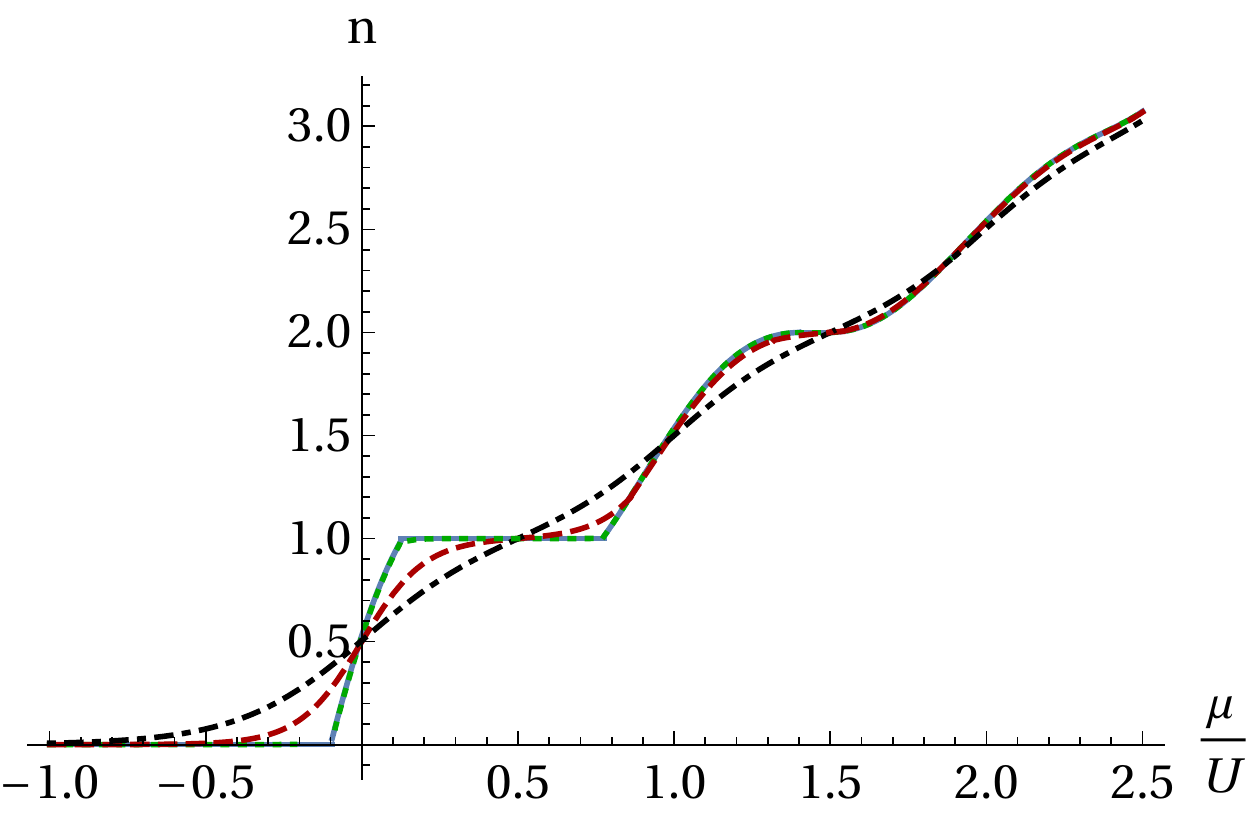} 
			\caption{}
			\label{pd2} \end{subfigure} \caption{(Color online)
			Equation of state for the hopping strengths (a) $tz/U=0.05$
			and (b) $tz/U=0.1$ and the
			temperatures $T=0$ (continuous blue), $\beta=30/U$
			(dotted green), $\beta=10/U$ (dashed red), and
			$\beta=5/U$ (dotted-dashed black).} \label{partdens}
	\end{figure}

We calculate the particle density, \begin{equation} n = -\frac{\partial
\mathcal{F}}{\partial \mu}, \end{equation} making use of our developed FTDPT.
We consider different temperatures and different hopping values for the purpose
of analyzing their effects on the density of particles. We plot the resulting
equation of state for two different values of the hopping parameter and four different values of the temperature,
thus observing the melting of the structure as in Refs.~\cite{bloch2,gerbier2}, as
shown in Fig. \ref{partdens}.

We observe the effects that the change of both the temperature and the hopping have upon the
particle density in Fig. \ref{partdens}. First, we conclude that increasing
the temperature makes the particle density to vary more smoothly when compared
to those particle densities with lower temperatures. This fact is due to
thermal fluctuations, which make the system more feasible to exist in the
superfluid phase. Also, by comparing the left panel to the right one we observe
the melting of the Mott lobes due to an increased hopping, which is also very
intuitive: the particles, having more kinetic energy, are more likely to hop
from one site to another, which is characteristic for the SF phase. Another factor responsible for making the curves smoother is the increase of the
chemical potential, $\mu/U$. The reason for this relies on the fact that the
bigger $\mu/U$ becomes the smaller the Mott lobes are, as can be seen in Fig.
\ref{pb_manyT}. Thus, the system is more likely to exist in the superfluid
phase for bigger values of $\mu/U$.

Now we must turn our attention to the points of the figures where the
degeneracies happen, which correspond to $\mu/U=1$ and $2$. We observe
that our calculations lead to no decreasing behavior happening at those regions, meaning that
our developed FTDPT method possess no inconsistency in the calculation of the
equation of state for the mean-field approximation of bosonic atoms confined in
optical lattices. As the NDPT leads to a vanishing behavior of the condensate densities, or at least a
decreasing one, in the vicinity of the degeneracies, i.e., $\mu/U=1$ and
$2$ in Fig. \ref{partdens}, we should also expect the same behavior in the
evaluation of the particle density. Finally, we also conclude that FTDPT gives
reliable results for the particle density since there is no decreasing behavior
in the vicinity of the degeneracies in Fig. \ref{partdens}.

\section{\label{sec4}Conclusions} By using a projection operator formalism we
were able to generalize the usual nondegenerate perturbation theory for the
mean-field approximation of the Bose-Hubbard Hamiltonian at finite
temperatures. With this we have solved the degeneracy problems which are
typical for nondegenerate perturbation theories. We
introduced the mean-field considerations applied to the Bose-Hubbard
Hamiltonian, followed by the Landau theory regarding second-order phase
transitions. Also, we showed that NDPT results in
an inconsistent behavior for the order parameter: it predicts a phase boundary
in a region of the phase diagram where there should be none. Subsequently, we developed a degenerate perturbative method based on a projection
operator formalism that corrects all such contradictions which arise from NDPT
due to degeneracies that occur between two adjacent Mott lobes. Our approach
was able to eliminate all the problematic divergences in the thermodynamic
potential, which allowed us to accurately evaluate the condensate densities and the particle densities in the vicinity of
the MI-SF phase transition for different temperatures and different hopping
values. Further, we drew the finite-temperature phase diagrams in order to
check the consistency of the previously calculated condensate densities.

It must be noted that for the zero-temperature regime, which is depicted in Fig.
\ref{cd_zeroT}, the results for the condensate densities are similar to those
obtained in \cite{martin}, which uses a Brillouin-Wigner treatment for the
perturbation expansion followed by a proper diagonalization in order to
calculate the system free energy. That approach differs from the one used in
this work. While the Brillouin-Wigner approach in \cite{martin} is also able
to correct the degeneracy problems from NDPT, it can only be applied to the
zero-temperature case. On the other hand, the theory presented in this paper
corrects degeneracy problems for both zero and finite temperatures, thus providing
a relatively simple method for calculating the condensate density in a wide range
of optical-lattice systems.

\section{Acknowledgments} We acknowledge financial support from the German
Research Foundation (DFG, Deutsche Forschungsgemeinschaft) within the Collaborative Research Center SFB/TR 49
"Condensed Matter Systems with Variable Many-Body Interactions" and SFB/TR 185
"Open System Control of Atomic and Photonic Matter" (OSCAR Project No.
277625399), and from the binational project between CAPES (Coordenação de Aperfeiçoamento de Pessoal de Nível Superior, Improvement Coordination of
Higher Level Personnel) and DAAD (Deutscher Akademischer Austauschdienst, German Academic Exchange Service) Probral (Programa Brasil-Alemanha, Brazil-Germany Program) No. 488/2018
Grant No. 88881.143936/2017-01. F.T.S. acknowledges CAPES for financial
support. We also thank M. Bonkhoff, S. Eggert, M. K\"ubler, and C. S\'a de Melo
for helpful discussions. Support from Optics and Photonics Research Center (CePOF, Centro de Pesquisas em Óptica e Fotônica) through Grant No. 2013/07276-1 is acknowledged.
F.E.A.d.S. acknowledges CNPq (Conselho Nacional de Denvolvimento Científico e Tecnológico, National Council for Scientific and Technological Development) for support through Grant No. 305586/2017-3.

\begin{appendix}
	
\section{\label{appendix1}Nondegenerate calculation of $\mathcal{Z}^{(2)}$ and
	$\mathcal{Z}^{(4)}$}

In this appendix we are concerned with the detailed calculation of
	$\mathcal{Z}^{(2)}$ and $\mathcal{Z}^{(4)}$ via the nondegenerate
	approach. The second-order term reads \begin{equation}\label{a1}
		\mathcal{Z}^{(2)}=\sum_{n=0}^{\infty}\mathrm{e}^{-\beta
		E_{n}}\int_{0}^{\beta}d\tau_{1}\int_{0}^{\tau_{1}}d\tau_{2}\langle
		n|\hat{V}_{\rm{I}}(\tau_{1})\hat{V}_{\rm{I}}(\tau_{2})|n\rangle.
	\end{equation} Inserting (\ref{pertua}) in (\ref{a1}) we have
	\begin{equation}\label{A2}
		\mathcal{Z}^{(2)}=\sum_{n=0}^{\infty}\mathrm{e}^{-\beta
		E_{n}}\int_{0}^{\beta}d\tau_{1}\int_{0}^{\tau_{1}}d\tau_{2}\langle
		n|\mathrm{e}^{\tau_{1}\hat{H}_{0}}\hat{V}\mathrm{e}^{-\tau_{1}\hat{H}_{0}}\mathrm{e}^{\tau_{2}\hat{H}_{0}}\hat{V}\mathrm{e}^{-\tau_{2}\hat{H}_{0}}|n\rangle.
	\end{equation} As $\ket{n}$ are eigenstates of $\hat{H}_0$ (\ref{A2})
	reduces to \begin{equation}
		\mathcal{Z}^{(2)}=\sum_{n=0}^{\infty}\mathrm{e}^{-\beta
		E_{n}}\int_{0}^{\beta}d\tau_{1}\int_{0}^{\tau_{1}}d\tau_{2}\mathrm{e}^{(\tau_{1}-\tau_{2})E_{n}}\langle
		n|\hat{V}\mathrm{e}^{-\tau_{1}\hat{H}_{0}}\mathrm{e}^{\tau_{2}\hat{H}_{0}}\hat{V}|n\rangle.
	\end{equation} According to (\ref{perturbation}) we have
	\begin{equation}
		\mathcal{Z}^{(2)}=t^{2}z^{2}\sum_{n=0}^{\infty}\mathrm{e}^{-\beta
		E_{n}}\int_{0}^{\beta}d\tau_{1}\int_{0}^{\tau_{1}}d\tau_{2}\mathrm{e}^{(\tau_{1}-\tau_{2})E_{n}}\langle
		n|\left(\Psi^{*}\hat{a}+\Psi
		\hat{a}^{\dagger}\right)\mathrm{e}^{-\tau_{1}\hat{H}_{0}}\mathrm{e}^{\tau_{2}\hat{H}_{0}}\left(\Psi^{*}\hat{a}+\Psi
		\hat{a}^{\dagger}\right)|n\rangle, \end{equation} yielding
		\begin{align}\label{A5}
			\mathcal{Z}^{(2)}&=&t^{2}z^{2}\sum_{n=0}^{\infty}\mathrm{e}^{-\beta
			E_{n}}\int_{0}^{\beta}d\tau_{1}\int_{0}^{\tau_{1}}d\tau_{2}\mathrm{e}^{(\tau_{1}-\tau_{2})E_{n}}\left(\Psi\sqrt{n}\langle
			n-1|+\Psi^{*}\sqrt{n+1}\langle n+1|\right) \nonumber \\
		&&\times\left(\Psi^{*}\sqrt{n}\mathrm{e}^{(\tau_{2}-\tau_{1})E_{n-1}}|n-1\rangle+\Psi\sqrt{n+1}\mathrm{e}^{(\tau_{2}-\tau_{1})E_{n+1}}|n+1\rangle\right).
		\end{align} The scalar products reduce (\ref{A5}) to
		\begin{equation}
			\mathcal{Z}^{(2)}=t^{2}z^{2}|\Psi|^{2}\sum_{n=0}^{\infty}\mathrm{e}^{-\beta
			E_{n}}\int_{0}^{\beta}d\tau_{1}\int_{0}^{\tau_{1}}d\tau_{2}\left(n\mathrm{e}^{(\tau_{1}-\tau_{2})\Delta_{n,n-1}}+(n+1)\mathrm{e}^{(\tau_{1}-\tau_{2})\Delta_{n,n+1}}\right).
		\end{equation} Finally, the integrations yield \begin{equation}
			\mathcal{Z}^{(2)}=t^{2}z^{2}|\Psi|^{2}\sum_{n=0}^{\infty}\mathrm{e}^{-\beta
			E_{n}}\left[n\left(\frac{\mathrm{e}^{\beta\Delta_{n,n-1}}-1}{\Delta_{n,n-1}^{2}}-\frac{\beta}{\Delta_{n,n-1}}\right)+(n+1)\left(\frac{\mathrm{e}^{\beta\Delta_{n,n+1}}-1}{\Delta_{n,n+1}^{2}}-\frac{\beta}{\Delta_{n,n+1}}\right)\right],
		\end{equation} where we have used the abbreviation
		$\Delta_{i,j}\equiv E_i-E_j$ for differences between two
		eigenvalues (\ref{En}).

For the fourth-order term we have \begin{equation}\label{A8}
	\mathcal{Z}^{(4)}=\sum_{n=0}^{\infty}\mathrm{e}^{-\beta
	E_{n}}\int_{0}^{\beta}d\tau_{1}\int_{0}^{\tau_{1}}d\tau_{2}\int_{0}^{\tau_{2}}d\tau_{3}\int_{0}^{\tau_{3}}d\tau_{4}\langle
	n|\hat{V}_{\rm{I}}(\tau_{1})\hat{V}_{\rm{I}}(\tau_{2})\hat{V}_{\rm{I}}(\tau_{3})\hat{V}_{\rm{I}}(\tau_{4})|n\rangle.
\end{equation} Inserting (\ref{H0}) and (\ref{pertua}) in (\ref{A8}) gives
	\begin{equation}\label{A9}
		\mathcal{Z}^{(4)}=\sum_{n=0}^{\infty}\mathrm{e}^{-\beta
		E_{n}}\int_{0}^{\beta}d\tau_{1}\int_{0}^{\tau_{1}}d\tau_{2}\int_{0}^{\tau_{2}}d\tau_{3}\int_{0}^{\tau_{3}}d\tau_{4}\mathrm{e}^{(\tau_{1}-\tau_{4})E_{n}}\langle
		n|\hat{V}\mathrm{e}^{-\tau_{1}\hat{H}_{0}}\hat{V}_{\rm{I}}(\tau_{2})\hat{V}_{\rm{I}}(\tau_{3})\mathrm{e}^{\tau_{4}\hat{H}_{0}}\hat{V}|n\rangle.
	\end{equation} According to (\ref{perturbation}) we have
	\begin{align}\label{A10}
		\mathcal{Z}^{(4)}&=t^{2}z^{2}\sum_{n=0}^{\infty}\mathrm{e}^{-\beta
		E_{n}}\int_{0}^{\beta}d\tau_{1}\int_{0}^{\tau_{1}}d\tau_{2}\int_{0}^{\tau_{2}}d\tau_{3}\int_{0}^{\tau_{3}}d\tau_{4}\mathrm{e}^{(\tau_{1}-\tau_{4})E_{n}}\left(\Psi\sqrt{n}\mathrm{e}^{(\tau_{2}-\tau_{1})E_{n-1}}\langle
		n-1|\right.\nonumber \\ &\left.
		+\Psi^{*}\sqrt{n+1}\mathrm{e}^{(\tau_{2}-\tau_{1})E_{n+1}}\langle
		n+1|\right)\hat{V}\mathrm{e}^{-\tau_{2}\hat{H}_{0}}\mathrm{e}^{\tau_{3}\hat{H}_{0}}\hat{V}\left(\Psi^{*}\sqrt{n}\mathrm{e}^{(\tau_{4}-\tau_{3})E_{n-1}}|n-1\rangle+\Psi\sqrt{n+1}\mathrm{e}^{(\tau_{4}-\tau_{3})E_{n+1}}|n+1\rangle\right).
	\end{align} Using again (\ref{H0}) and (\ref{pertua}) in (\ref{A10})
	results in \begin{align}
		\mathcal{Z}^{(4)}&=t^{4}z^{4}\sum_{n=0}^{\infty}\mathrm{e}^{-\beta
		E_{n}}\int_{0}^{\beta}d\tau_{1}\int_{0}^{\tau_{1}}d\tau_{2}\int_{0}^{\tau_{2}}d\tau_{3}\int_{0}^{\tau_{3}}d\tau_{4}\mathrm{e}^{(\tau_{1}-\tau_{4})E_{n}}\nonumber
		\\
		&\times\left[\Psi\sqrt{n}\mathrm{e}^{(\tau_{2}-\tau_{1})E_{n-1}}\left(\Psi\sqrt{n-1}\mathrm{e}^{-\tau_{2}E_{n-2}}\langle
		n-2|+\Psi^{*}\sqrt{n}\mathrm{e}^{-\tau_{2}E_{n}}\langle
		n|\right)\right.\nonumber \\ &\left.
		+\Psi^{*}\sqrt{n+1}\mathrm{e}^{(\tau_{2}-\tau_{1})E_{n+1}}\left(\Psi\sqrt{n+1}\mathrm{e}^{-\tau_{2}E_{n}}\langle
		n|+\Psi^{*}\sqrt{n+2}\mathrm{e}^{-\tau_{2}E_{n+2}}\langle
		n+2|\right)\right] \nonumber \\
		&\times\left[\Psi^{*}\sqrt{n}\mathrm{e}^{(\tau_{4}-\tau_{3})E_{n-1}}\left(\Psi^{*}\sqrt{n-1}\mathrm{e}^{\tau_{3}E_{n-2}}|n-2\rangle+\Psi\sqrt{n}\mathrm{e}^{\tau_{3}E_{n}}|n\rangle\right)\right.\nonumber
		\\ &\left.
	+\Psi\sqrt{n+1}\mathrm{e}^{(\tau_{4}-\tau_{3})E_{n+1}}\left(\Psi^{*}\sqrt{n+1}\mathrm{e}^{\tau_{3}E_{n}}|n\rangle+\Psi\sqrt{n+2}\mathrm{e}^{\tau_{3}E_{n+2}}|n+2\rangle\right)\right],
	\end{align} which, from the scalar products, reduces to \begin{align}
		\mathcal{Z}^{(4)}&=t^{4}z^{4}\big|\Psi\big|^{4}\sum_{n=0}^{\infty}\mathrm{e}^{-\beta
		E_{n}}\int_{0}^{\beta}d\tau_{1}\int_{0}^{\tau_{1}}d\tau_{2}\int_{0}^{\tau_{2}}d\tau_{3}\int_{0}^{\tau_{3}}d\tau_{4}\nonumber
		\\ &\times\left[
			n(n-1)\mathrm{e}^{(\tau_{1}-\tau_{4})\Delta_{n,n-1}}\mathrm{e}^{(\tau_{2}-\tau_{3})\Delta_{n-1,n-2}}+(n+1)(n+2)\mathrm{e}^{(\tau_{1}-\tau_{4})\Delta_{n,n+1}}\mathrm{e}^{(\tau_{2}-\tau_{3})\Delta_{n+1,n+2}}\right.\nonumber
			\\
			&+n^{2}\mathrm{e}^{(\tau_{1}-\tau_{4})\Delta_{n,n-1}}\mathrm{e}^{(\tau_{2}-\tau_{3})\Delta_{n-1,n}}+n(n+1)\mathrm{e}^{(\tau_{1}-\tau_{2})\Delta_{n,n-1}}\mathrm{e}^{(\tau_{3}-\tau_{4})\Delta_{n,n+1}}\nonumber
			\\
			&\left.+n(n+1)\mathrm{e}^{(\tau_{1}-\tau_{2})\Delta_{n,n+1}}\mathrm{e}^{(\tau_{3}-\tau_{4})\Delta_{n,n-1}}+(n+1)^{2}\mathrm{e}^{(\tau_{1}-\tau_{2})\Delta_{n,n+1}}\mathrm{e}^{(\tau_{3}-\tau_{4})\Delta_{n,n+1}}\right].
	\end{align}

The integrations result in \begin{align}
	\mathcal{Z}^{(4)}&=t^{4}z^{4}\big|\Psi\big|^{4}\sum_{n=0}^{\infty}\mathrm{e}^{-\beta
	E_{n}}\left\{
		n\left(n-1\right)\frac{\mathrm{e}^{\beta\Delta_{n,n-2}}-1}{\Delta_{n,n-1}\Delta_{n-1,n-2}\Delta_{n,n-2}}\left(\frac{1}{\Delta_{n-1,n-2}}-\frac{1}{\Delta_{n,n-2}}\right)\right.
		\nonumber\\ &\left.
		+n\left(n-1\right)\frac{\mathrm{e}^{\beta\Delta_{n,n-1}}-1}{\Delta_{n,n-1}^{2}\Delta_{n,n-2}}\left(\frac{1}{\Delta_{n,n-1}}+\frac{1}{\Delta_{n-1,n-2}}\right)
		\right. \nonumber\\
		&+n\left(n-1\right)\frac{\mathrm{e}^{\beta\Delta_{n,n-1}}-1}{\Delta_{n,n-1}^{2}\Delta_{n-1,n-2}}\left(\frac{1}{\Delta_{n,n-1}}-\frac{1}{\Delta_{n-1,n-2}}\right)-n\left(n-1\right)\frac{\beta}{\Delta_{n,n-1}^{2}}\left(\frac{\mathrm{e}^{\beta\Delta_{n,n-1}}}{\Delta_{n-1,n-2}}+\frac{1}{\Delta_{n,n-2}}\right)\nonumber\\
		&+\left(n+1\right)\left(n+2\right)\frac{\mathrm{e}^{\beta\Delta_{n,n+2}}-1}{\Delta_{n,n+1}\Delta_{n+1,n+2}\Delta_{n,n+2}}\left(\frac{1}{\Delta_{n+1,n+2}}-\frac{1}{\Delta_{n,n+2}}\right)\nonumber\\
		&+\left(n+1\right)\left(n+2\right)\frac{\mathrm{e}^{\beta\Delta_{n,n+1}}-1}{\Delta_{n,n+1}^{2}\Delta_{n,n+2}}\left(\frac{1}{\Delta_{n,n+1}}+\frac{1}{\Delta_{n+1,n+2}}\right)\nonumber\\
		&+\left(n+1\right)\left(n+2\right)\frac{\mathrm{e}^{\beta\Delta_{n,n+1}}-1}{\Delta_{n,n+1}^{2}\Delta_{n+1,n+2}}\left(\frac{1}{\Delta_{n,n+1}}-\frac{1}{\Delta_{n+1,n+2}}\right)-\left(n+1\right)\left(n+2\right)\frac{\beta}{\Delta_{n,n+1}^{2}}\left(\frac{\mathrm{e}^{\beta\Delta_{n,n+1}}}{\Delta_{n+1,n+2}}+\frac{1}{\Delta_{n,n+2}}\right)\nonumber\\
		&+3n^{2}\frac{1-\mathrm{e}^{\beta\Delta_{n,n-1}}}{\Delta_{n,n-1}^{4}}+n^{2}\frac{\beta}{\Delta_{n,n-1}^{3}}\left(2+\mathrm{e}^{\beta\Delta_{n,n-1}}\right)+n^{2}\frac{\beta^{2}}{2\Delta_{n,n-1}^{2}}\nonumber\\
		&+\frac{n\left(n+1\right)}{\Delta_{n,n+1}^{2}\Delta_{n-1,n+1}}\left(\frac{\mathrm{e}^{\beta\Delta_{n,n+1}}-1}{\Delta_{n,n+1}}+\frac{1-\mathrm{e}^{\beta\Delta_{n,n-1}}}{\Delta_{n,n-1}}\right)+n\left(n+1\right)\frac{1-\mathrm{e}^{\beta\Delta_{n,n-1}}}{\Delta_{n,n-1}^{2}\Delta_{n,n+1}}\left(\frac{1}{\Delta_{n,n-1}}+\frac{1}{\Delta_{n,n+1}}\right)\nonumber\\
		&+n\left(n+1\right)\frac{\beta}{\Delta_{n,n-1}\Delta_{n,n+1}}\left(\frac{1}{\Delta_{n,n-1}}+\frac{1}{\Delta_{n,n+1}}\right)+n\left(n+1\right)\frac{\beta^{2}}{2\Delta_{n,n-1}\Delta_{n,n+1}}\nonumber\\
		&+\frac{n\left(n+1\right)}{\Delta_{n,n-1}^{2}\Delta_{n+1,n-1}}\left(\frac{\mathrm{e}^{\beta\Delta_{n,n-1}}-1}{\Delta_{n,n-1}}+\frac{1-\mathrm{e}^{\beta\Delta_{n,n+1}}}{\Delta_{n,n+1}}\right)+n\left(n+1\right)\frac{1-\mathrm{e}^{\beta\Delta_{n,n+1}}}{\Delta_{n,n+1}^{2}\Delta_{n,n-1}}\left(\frac{1}{\Delta_{n,n+1}}+\frac{1}{\Delta_{n,n-1}}\right)\nonumber\\
		&+n\left(n+1\right)\frac{\beta}{\Delta_{n,n+1}\Delta_{n,n-1}}\left(\frac{1}{\Delta_{n,n+1}}+\frac{1}{\Delta_{n,n-1}}\right)+n\left(n+1\right)\frac{\beta^{2}}{2\Delta_{n,n+1}\Delta_{n,n-1}}\nonumber\\
		&\left.+3\left(n+1\right)^{2}\frac{1-\mathrm{e}^{\beta\Delta_{n,n+1}}}{\Delta_{n,n+1}^{4}}+\left(n+1\right)^{2}\frac{\beta}{\Delta_{n,n+1}^{3}}\left(2+\mathrm{e}^{\beta\Delta_{n,n+1}}\right)+\left(n+1\right)^{2}\frac{\beta^{2}}{2\Delta_{n,n+1}^{2}}\right\}.
\end{align}

\section{\label{appendix}Degenerate calculation of $\mathcal{Z}^{(2)}$}
	
This appendix is devoted to the evaluation of (\ref{Z}) for the second order of
(\ref{U}) \begin{align} \mathcal{Z}^{(2)} &=& \expo{-\beta \mathcal{E}_+}
	\int_{0}^{\beta} d\tau_1 \int_{0}^{\tau_1} d\tau_2 \, \bra{\Phi_+}
	\hat{\mathcal{V}}_{\rm{I}}(\tau_1) \hat{\mathcal{V}}_{\rm{I}}(\tau_2)
	\ket{\Phi_+} + \expo{-\beta \mathcal{E}_-} \int_{0}^{\beta} d\tau_1
	\int_{0}^{\tau_1} d\tau_2 \, \bra{\Phi_-}
	\hat{\mathcal{V}}_{\rm{I}}(\tau_1) \hat{\mathcal{V}}_{\rm{I}}(\tau_2)
	\ket{\Phi_-} \nonumber \\ &&+ \sum_{m \in Q} \expo{-\beta E_m}
\int_{0}^{\beta} d\tau_1 \int_{0}^{\tau_1} d\tau_2 \, \bra{m}
\hat{\mathcal{V}}_{\rm{I}}(\tau_1) \hat{\mathcal{V}}_{\rm{I}}(\tau_2) \ket{m}.
\end{align} We shall perform the calculation of each term separately and
identify them as $\mathcal{Z}^{(2)}=\mathcal{Z}^{(2)}_+ + \mathcal{Z}^{(2)}_- +
\mathcal{Z}^{(2)}_m$.

As the evaluation of $\mathcal{Z}^{(2)}_+$ and $\mathcal{Z}^{(2)}_-$ are
completely equivalent we perform a generic calculation for both contributions.
Inserting the expression for the perturbation in the interaction picture
(\ref{pertu}) in the first term we have \begin{align} \mathcal{Z}^{(2)}_\pm =
	\expo{-\beta \mathcal{E}_\pm} \int_{0}^{\beta} d\tau_1
	\int_{0}^{\tau_1} d\tau_2 \, \bra{\Phi_\pm} \expo{\tau_1
	\hat{\mathcal{H}}_{0}}
	\left(\hat{P}\hat{V}\hat{Q}+\hat{Q}\hat{V}\hat{P}+\hat{Q}\hat{V}\hat{Q}\right)
	\expo{-\tau_1 \hat{\mathcal{H}}_{0}} \expo{\tau_2
	\hat{\mathcal{H}}_{0}}
	\left(\hat{P}\hat{V}\hat{Q}+\hat{Q}\hat{V}\hat{P}+\hat{Q}\hat{V}\hat{Q}\right)
	\expo{-\tau_2 \hat{\mathcal{H}}_{0}} \ket{\Phi_\pm}.  \end{align} As
	$\ket{\Phi_\pm}$ are eigenstates of $\hat{\mathcal{H}}_0$ we get
	\begin{align}\label{b3} \mathcal{Z}^{(2)}_\pm = \expo{-\beta
		\mathcal{E}_\pm} \int_{0}^{\beta} d\tau_1 \int_{0}^{\tau_1}
		d\tau_2 \, \expo{(\tau_1-\tau_2) \mathcal{E}_\pm}
		\bra{\Phi_\pm}
		\left(\hat{P}\hat{V}\hat{Q}+\hat{Q}\hat{V}\hat{P}+\hat{Q}\hat{V}\hat{Q}\right)
		\expo{-\tau_1 \hat{\mathcal{H}}_{0}} \expo{\tau_2
		\hat{\mathcal{H}}_{0}}
		\left(\hat{P}\hat{V}\hat{Q}+\hat{Q}\hat{V}\hat{P}+\hat{Q}\hat{V}\hat{Q}\right)
		\ket{\Phi_\pm}.  \end{align} As we have $\hat{Q} \ket{\Phi_\pm}
		= 0$, $\hat{P} \ket{\Phi_\pm} = \ket{\Phi_\pm}$ as well as
		$\hat{Q}$ and $\hat{P}$ represent hermitian operators
		(\ref{b3}) reduces to \begin{align} \mathcal{Z}^{(2)}_\pm =
			\expo{-\beta \mathcal{E}_\pm} \int_{0}^{\beta} d\tau_1
			\int_{0}^{\tau_1} d\tau_2 \, \expo{(\tau_1-\tau_2)
			\mathcal{E}_\pm} \bra{\Phi_\pm}  \hat{V}\hat{Q}
			\expo{-\tau_1 \hat{\mathcal{H}}_{0}} \expo{\tau_2
			\hat{\mathcal{H}}_{0}} \hat{Q}\hat{V} \ket{\Phi_\pm}.
		\end{align} From (\ref{perturbation}) and (\ref{PHI+-}) and
		using the scalar products \begin{subequations} \begin{align}
			\braket{n}{\Phi_\pm} &=
			\left(1+\frac{\left|\mathcal{E}_{\pm}-E_{n}\right|^{2}}{t^2
			z^2\left|\Psi\right|^{2}\left(n+1\right)}\right)^{-1/2}, \\
			\braket{n+1}{\Phi_\pm} &=
			\left(1+\frac{\left|\mathcal{E}_{\pm}-E_{n}\right|^{2}}{t^2
			z^2\left|\Psi\right|^{2}\left(n+1\right)}\right)^{-1/2}
			\frac{E_{n}-\mathcal{E}_{\pm}}{t
			z\sqrt{\left|\Psi\right|^{2}\left(n+1\right)}},
		\end{align} \end{subequations} we have \begin{align}\label{**}
			\mathcal{Z}^{(2)}_\pm &=& \expo{-\beta \mathcal{E}_\pm}
			\int_{0}^{\beta} d\tau_1 \int_{0}^{\tau_1} d\tau_2 \,
			\expo{(\tau_1-\tau_2) \mathcal{E}_\pm} t^2 z^2
			\left(\Psi \braket{\Phi_\pm}{n} \sqrt{n} \bra{n-1} +
			\Psi^* \braket{\Phi_\pm}{n+1}\sqrt{n+2}\bra{n+2}\right)
			\nonumber \\ &&\times\expo{-\tau_1
		\hat{\mathcal{H}}_{0}} \expo{\tau_2 \hat{\mathcal{H}}_{0}}
		\left(\Psi^{*} \braket{n}{\Phi_\pm} \sqrt{n} \ket{n-1} + \Psi
		\braket{n+1}{\Phi_\pm}\sqrt{n+2}\ket{n+2}\right).  \end{align}
		The evaluation of (\ref{**}) leads to \begin{align}
			\mathcal{Z}^{(2)}_\pm &= t^2 z^2 \big|\Psi\big|^2
			\expo{-\beta \mathcal{E}_\pm} \int_{0}^{\beta} d\tau_1
			\int_{0}^{\tau_1} d\tau_2 \,
			\left(\expo{(\tau_1-\tau_2)\Delta_{\pm,n-1}} n
			\big|\braket{\Phi_\pm}{n}\big|^2 +
			\expo{(\tau_1-\tau_2)\Delta_{\pm,n+2}} (n+2)
			\big|\braket{\Phi_\pm}{n+1}\big|^2 \right). \label{B5}
		\end{align}

Evaluating the integrations in (\ref{B5}) yields finally
\begin{align}\label{Z+-} \mathcal{Z}^{(2)}_\pm &= t^2 z^2 \big|\Psi\big|^2
	\expo{-\beta \mathcal{E}_\pm} \left[ n \big|\braket{\Phi_\pm}{n}\big|^2
	\left(\frac{\expo{\beta
	\Delta_{\pm,n-1}}-1}{\Delta_{\pm,n-1}^2}-\frac{\beta}{\Delta_{\pm,n-1}}\right)+
	(n+2) \big|\braket{\Phi_\pm}{n+1}\big|^2 \left(\frac{\expo{\beta
	\Delta_{\pm,n+2}}-1}{\Delta_{\pm,n+2}^2}-\frac{\beta}{\Delta_{\pm,n+2}}\right)\right].
\end{align}

The last term to be calculated is $\mathcal{Z}_m^{(2)}$. The first steps of
this calculation are similar to those from the evaluation of
$\mathcal{Z}_{\pm}^{(2)}$. Therefore, we have \begin{align} \mathcal{Z}_m^{(2)}
	&= \sum_{m \in Q} \expo{-\beta E_m} \int_{0}^{\beta} d\tau_1
	\int_{0}^{\tau_1} d\tau_2 \,  \bra{m} \expo{\tau_1
	\hat{\mathcal{H}}_{0}}
	\left(\hat{P}\hat{V}\hat{Q}+\hat{Q}\hat{V}\hat{P}+\hat{Q}\hat{V}\hat{Q}\right)
	\expo{-\tau_1 \hat{\mathcal{H}}_{0}} \expo{\tau_2
	\hat{\mathcal{H}}_{0}}
	\left(\hat{P}\hat{V}\hat{Q}+\hat{Q}\hat{V}\hat{P}+\hat{Q}\hat{V}\hat{Q}\right)
	\expo{-\tau_2 \hat{\mathcal{H}}_{0}} \ket{m} \nonumber \\ &= \sum_{m
	\in Q} \expo{-\beta E_m} \int_{0}^{\beta} d\tau_1 \int_{0}^{\tau_1}
	d\tau_2 \, \expo{(\tau_1-\tau_2)E_m} \bra{m} \hat{V} \expo{-\tau_1
	\hat{\mathcal{H}}_{0}} \expo{\tau_2 \hat{\mathcal{H}}_{0}}
	\hat{V}\ket{m} \nonumber \\ &= t^2 z^2 \sum_{m \in Q} \expo{-\beta E_m}
	\int_{0}^{\beta} d\tau_1 \int_{0}^{\tau_1} d\tau_2 \,
	\expo{(\tau_1-\tau_2)E_m} \left(\Psi \sqrt{m} \bra{m-1} + \Psi^*
	\sqrt{m+1} \bra{m+1} \right)  \nonumber \\ &\times\expo{-\tau_1
	\hat{\mathcal{H}}_{0}} \expo{\tau_2 \hat{\mathcal{H}}_{0}} \left(\Psi^*
	\sqrt{m} \ket{m-1} + \Psi \sqrt{m+1} \ket{m+1} \right).  \end{align}
	Applying the exponential operators to the eigenstates we are left with
	\begin{align} \mathcal{Z}_m^{(2)}&= t^2 z^2 \sum_{m \in Q} \expo{-\beta
		E_m} \int_{0}^{\beta} d\tau_1 \int_{0}^{\tau_1} d\tau_2 \,
		\expo{(\tau_1-\tau_2)E_m}  \nonumber \\ &\times\left[\Psi
		\sqrt{m} \left(\expo{-\tau_1 \mathcal{E}_+}
		\braket{m-1}{\Phi_+} \bra{\Phi_+} + \expo{-\tau_1
		\mathcal{E}_-} \braket{m-1}{\Phi_-} \bra{\Phi_-} + \sum_{m' \in
		Q} \expo{-\tau_1 E_{m'}} \braket{m-1}{m'} \bra{m'} \right)
		\right. \nonumber \\ & \left. + \Psi^* \sqrt{m+1}
		\left(\expo{-\tau_1 \mathcal{E}_+} \braket{m+1}{\Phi_+}
		\bra{\Phi_+} + \expo{-\tau_1 \mathcal{E}_-}
		\braket{m+1}{\Phi_-} \bra{\Phi_-} + \sum_{m'' \in Q}
		\expo{-\tau_1 E_{m''}} \braket{m+1}{m''} \bra{m''}
		\right)\right]  \nonumber \\ &\times \left[\Psi^* \sqrt{m}
		\left(\expo{\tau_2 \mathcal{E}_+} \braket{\Phi_+}{m-1}
		\ket{\Phi_+} + \expo{\tau_2 \mathcal{E}_-} \braket{\Phi_-}{m-1}
		\ket{\Phi_-} + \sum_{m'''\in Q} \expo{\tau_2 E_{m'''}}
		\braket{m'''}{m-1} \ket{m'''} \right) \right. \nonumber \\ &
		\left. + \Psi \sqrt{m+1} \left(\expo{\tau_2 \mathcal{E}_+}
		\braket{\Phi_+}{m+1} \ket{\Phi_+} + \expo{\tau_2 \mathcal{E}_-}
		\braket{\Phi_-}{m+1} \ket{\Phi_-} + \sum_{m''''\in Q}
		\expo{\tau_2 E_{m''''}} \braket{m''''}{m+1} \ket{m''''} \right)
		\right].  \end{align} When we evaluate the multiplication among
		the terms between brackets, we must be aware of the fact that
		the cross terms, i.e., those that contain $\Psi^2$ or
		$\Psi^{*2}$ give zero since they contain the products
		$\braket{m-1}{\Phi_\pm}$ and $\braket{m+1}{\Phi_\pm}$, which
		cannot be both nonzero because it is not possible for $m+1$ and
		$m-1$ be equal to $n$ or $n+1$ at the same time. So, we are
		left with \begin{align} \mathcal{Z}_m^{(2)}&= t^2 z^2 |\Psi|^2
			\sum_{m \in Q} \expo{-\beta E_m} \int_{0}^{\beta}
			d\tau_1 \int_{0}^{\tau_1} d\tau_2 \, \Bigg[m
			\expo{(\tau_1-\tau_2) \Delta_{m,+}}
			\big|\braket{\Phi_+}{m-1}\big|^2 + m
			\expo{(\tau_1-\tau_2) \Delta_{m,-}}
			\big|\braket{\Phi_-}{m-1}\big|^2  \nonumber \\ &+ m
			\sum_{m'\in Q} \expo{(\tau_1-\tau_2) \Delta_{m,m'}}
			\big|\braket{m-1}{m'}\big|^2 + (m+1)
			\expo{(\tau_1-\tau_2) \Delta_{m,+}}
			\big|\braket{\Phi_+}{m+1}\big|^2  \nonumber \\ & +
			(m+1) \expo{(\tau_1-\tau_2) \Delta_{m,-}}
			\big|\braket{\Phi_-}{m+1}\big|^2 + (m+1) \sum_{m''\in
			Q} \expo{(\tau_1-\tau_2) \Delta_{m,m''}}
			\big|\braket{m+1}{m''}\big|^2 \Bigg] .  \end{align} The
			integrations lead finally to \begin{align}\label{Zm}
				\mathcal{Z}_m^{(2)}&= t^2 z^2 |\Psi|^2 \sum_{m
				\in Q} \expo{-\beta E_m} \Bigg[m
				\big|\braket{\Phi_+}{m-1}\big|^2 \left(\frac{
					\expo{\beta
					\Delta_{m,+}}-1}{\Delta_{m,+}^2} -
					\frac{\beta}{\Delta_{m,+}}\right) + m
					\big|\braket{\Phi_-}{m-1}\big|^2
					\left(\frac{ \expo{\beta
					\Delta_{m,-}}-1}{\Delta_{m,-}^2} -
					\frac{\beta}{\Delta_{m,-}}\right)
					\nonumber \\ &  + m \sum_{m'\in Q}
					\left(\frac{ \expo{\beta
					\Delta_{m,m'}}-1}{\Delta_{m,m'}^2} -
					\frac{\beta}{\Delta_{m,m'}} \right)
					\big|\braket{m-1}{m'}\big|^2 + (m+1)
					\big|\braket{\Phi_+}{m+1}\big|^2
					\left(\frac{ \expo{\beta
					\Delta_{m,+}}-1}{\Delta_{m,+}^2} -
					\frac{\beta}{\Delta_{m,+}}\right) \\ &
					+ (m+1)
					\big|\braket{\Phi_-}{m+1}\big|^2
					\left(\frac{ \expo{\beta
					\Delta_{m,-}}-1}{\Delta_{m,-}^2} -
					\frac{\beta}{\Delta_{m,-}}\right) +
					(m+1) \sum_{m''\in Q} \left(\frac{
						\expo{\beta
						\Delta_{m,m''}}-1}{\Delta_{m,m''}^2}
						-
						\frac{\beta}{\Delta_{m,m''}}\right)
						\big|\braket{m+1}{m''}\big|^2
						\Bigg] . \nonumber \end{align}

Combining the contributions (\ref{Z+-}) and (\ref{Zm}) the second-order term of
the partition function reads \begin{align} \mathcal{Z}^{(2)}&=t^2 z^2
	\big|\Psi\big|^2 \expo{-\beta \mathcal{E}_+} \left[ n
	\big|\braket{\Phi_+}{n}\big|^2 \left(\frac{\expo{\beta
	\Delta_{+,n-1}}-1}{\Delta_{+,n-1}^2}-\frac{\beta}{\Delta_{+,n-1}}\right)+
	(n+2) \big|\braket{\Phi_+}{n+1}\big|^2 \left(\frac{\expo{\beta
	\Delta_{+,n+2}}-1}{\Delta_{+,n+2}^2}-\frac{\beta}{\Delta_{+,n+2}}\right)\right]
	\nonumber \\ &+t^2 z^2 \big|\Psi\big|^2 \expo{-\beta \mathcal{E}_-}
	\left[ n \big|\braket{\Phi_-}{n}\big|^2 \left(\frac{\expo{\beta
	\Delta_{-,n-1}}-1}{\Delta_{-,n-1}^2}-\frac{\beta}{\Delta_{-,n-1}}\right)+
	(n+2) \big|\braket{\Phi_-}{n+1}\big|^2 \left(\frac{\expo{\beta
	\Delta_{-,n+2}}-1}{\Delta_{-,n+2}^2}-\frac{\beta}{\Delta_{-,n+2}}\right)\right]
	\nonumber \\ &+ t^2 z^2 \big|\Psi\big|^2 \sum_{m \in Q} \expo{-\beta
	E_m} \Bigg[m \big|\braket{\Phi_+}{m-1}\big|^2 \left(\frac{ \expo{\beta
	\Delta_{m,+}}-1}{\Delta_{m,+}^2} - \frac{\beta}{\Delta_{m,+}}\right) +
	m \big|\braket{\Phi_-}{m-1}\big|^2 \left(\frac{ \expo{\beta
	\Delta_{m,-}}-1}{\Delta_{m,-}^2} - \frac{\beta}{\Delta_{m,-}}\right)
	\nonumber \\ &  + m \sum_{m'\in Q} \left(\frac{ \expo{\beta
	\Delta_{m,m'}}-1}{\Delta_{m,m'}^2} - \frac{\beta}{\Delta_{m,m'}}
	\right) \big|\braket{m-1}{m'}\big|^2 + (m+1)
	\big|\braket{\Phi_+}{m+1}\big|^2 \left(\frac{ \expo{\beta
	\Delta_{m,+}}-1}{\Delta_{m,+}^2} - \frac{\beta}{\Delta_{m,+}}\right) \\
	& + (m+1) \big|\braket{\Phi_-}{m+1}\big|^2 \left(\frac{ \expo{\beta
	\Delta_{m,-}}-1}{\Delta_{m,-}^2} - \frac{\beta}{\Delta_{m,-}}\right) +
	(m+1) \sum_{m''\in Q} \left(\frac{ \expo{\beta
	\Delta_{m,m''}}-1}{\Delta_{m,m''}^2} -
	\frac{\beta}{\Delta_{m,m''}}\right) |\braket{m+1}{m''}|^2 \Bigg] .
	\nonumber \end{align}

Taking into account that the scalar products $\braket{m-1}{m'}$ and
$\braket{m+1}{m''}$ lead to one further restriction each in the summations,
thus we finally obtain equation (\ref{Z2}).

\end{appendix}

% % % % % % % % % % % % % % % % % % % % % % % % % % % % % % % % % % % % % % % %
% % % % % % % % % % % % % % % % % % % % % % % % % % % % % % % % % % % % % % % %
% % % % % % % % % % % % % % % % % % % % % % % % % % % % % % % % % % % % % % % %
% % % %

% % % % % % % % % % % % % % % % % % % % % % % % % % % % % % % % % % % % % % % %
% % % % % % % % % % % % % % % % % % % % % % % % % % % % % % % % % % % % % % % %
% % % % % % % % % % % % % % % % % % % % % % % % % % % % % % % % % % % % % % % %
% % % %

\bibliography{references}{}
\bibliographystyle{unsrt}%Used BibTeX style is

\end{document}